\documentclass{emulateapj}
\usepackage{lscape}

\newcommand{\kms}{km~s$^{-1}$}
\newcommand{\brg}{Br$\gamma$}

\newcommand{\HII}{H\,{\sc ii}}
\newcommand{\hei}{He\,{\sc i}}

\newcommand{\myemail}{abik@mpia.de}

\shorttitle{Sequential star formation in RCW 34}
\shortauthors{Bik et al.}

\begin{document}


\title{Sequential Star Formation in RCW 34: \\
A Spectroscopic Census of the Stellar Content of High-mass Star-forming Regions \altaffilmark{1} }


\author{A. Bik\altaffilmark{2},
E. Puga\altaffilmark{3,4},
L.B.F.M. Waters\altaffilmark{4,5},
M. Horrobin\altaffilmark{5,6},
Th. Henning\altaffilmark{2},
T. Vasyunina\altaffilmark{2},
H. Beuther\altaffilmark{2},
H. Linz\altaffilmark{2},
L. Kaper\altaffilmark{5},
M. van den Ancker\altaffilmark{7},
 A. Lenorzer\altaffilmark{8},
E. Churchwell\altaffilmark{9},
S. Kurtz\altaffilmark{10},
M. B. N. Kouwenhoven\altaffilmark{11,12} ,
A. Stolte\altaffilmark{6},
A. de Koter\altaffilmark{5,13},
 W. F. Thi\altaffilmark{14},
F. Comer\'{o}n\altaffilmark{7},
Ch. Waelkens \altaffilmark{4}
}

\email{\myemail}
\altaffiltext{1}{Based on observations collected at the European Southern Observatory at Paranal, Chile (ESO program 078.C-0780).}
\altaffiltext{2}{Max-Planck-Institut f\"ur Astronomie, K\"onigstuhl 17, 69117 Heidelberg, Germany}
\altaffiltext{3}{Centro de Astrobiolog{\'{i}}a (CSIC-INTA), 28850 Torrej\'{o}n de Ardoz, Madrid, Spain}
\altaffiltext{4}{Instituut voor Sterrenkunde, Celestijnenlaan 200D, B-3001 Leuven, Belgium}
\altaffiltext{5}{Sterrenkundig Instituut "Anton Pannekoek", University of Amsterdam, Science Park 904, 1098 XH Amsterdam, The Netherlands}
\altaffiltext{6}{I.Physikalisches Institut, Universit\"{a}t zu K\"{o}ln, 50937 K\"{o}ln, Germany}
\altaffiltext{7}{European Southern Observatory, Karl-Schwarzschild Strasse 2, Garching-bei-M\"{u}nchen, D-85748, Germany}
\altaffiltext{8}{Instituto Astrof\'{i}sica de Canarias, V\'ia L\'actea s/n, E-38200 La Laguna, Spain}
\altaffiltext{9}{University of Wisconsin, 475 North Charter Street, WI53706 Madison, USA}
\altaffiltext{10}{CRyA, Universidad Nacional Aut\'onoma de M\'exico, Apdo. 3-72, 58090 Morelia, Michoac\'an, M\'exico}
\altaffiltext{11}{The Kavli Institute for Astronomy and Astrophysics, Peking University, Yi He Yuan Lu 5, Hai Dian Qu, Beijing 100871, China}
\altaffiltext{12}{University of Sheffield, Hicks Building, Hounsfixeld Road, Sheffield S3 7RH, United Kingdom}
\altaffiltext{13}{Astronomical Institute, Utrecht University, Princetonplein 5, 3584CC Utrecht, The Netherlands}
\altaffiltext{14}{SUPA, Institute for Astronomy, University of Edinburgh, Royal Observatory Edinburgh, Backford Hill, Edinburgh, EH9 3HJ, United Kingdom}


\begin{abstract}

In this paper we present VLT/SINFONI integral field spectroscopy of RCW 34 along with \emph{Spitzer}/IRAC photometry of the surroundings. RCW 34 consists of three different regions. A large bubble has been detected on the IRAC images in which a cluster of intermediate- and low-mass class II objects is found. At the northern edge of this bubble, an \HII\ region  is located, ionized by 3 OB stars, of which the most massive star has spectral type O8.5V. Intermediate mass stars (2 - 3 M$_\sun$) are detected of G- and K- spectral type. These stars are still in the pre-main sequence (PMS)  phase.  North of the \HII\ region, a photon-dominated region is present, marking the edge of a dense molecular cloud traced by H$_{2}$ emission.  Several class 0/I objects are associated with this cloud, indicating that star formation is still taking place. 

The distance to RCW 34 is revised to 2.5 $\pm$ 0.2 kpc and an age estimate of 2 $\pm$ 1 Myrs is derived from the properties of the PMS stars inside the \HII\ region. Between the class II sources in the bubble and the  PMS stars in the \HII\ region, no age difference could be detected with the present data. The presence of class 0/I sources in the molecular cloud, however, suggests that the objects inside the molecular cloud are significantly younger.

The most likely scenario for the formation of the three regions is that star formation propagates from South to North. 
First the bubble is formed, produced by intermediate- and low-mass stars only, after that, the \HII\ region is formed from a dense core at the edge of the molecular cloud, resulting in the expansion as a champagne flow.  More recently, star formation occurred in the rest of the molecular cloud. Two different formation scenarios are possible: (a) The bubble with the cluster of low- and intermediate mass stars triggered the formation of the O star at the edge of the molecular cloud which in turn induces the current star-formation in the molecular cloud. (b) An external triggering is responsible for the star-formation propagating from South to North. 

\end{abstract}


\section{Introduction}

One of the key questions in astrophysics is understanding the formation and early evolution of high-mass stars.  They play a major role in shaping the interstellar medium due to their strong stellar winds and UV radiation fields. Observations show that high-mass stars preferentially form in clusters; therefore, if we wish to understand the formation of high-mass stars, we need to understand the clustered mode of star formation \citep{Zinnecker07}. 

The actual formation of massive stars happens deeply embedded in molecular clouds behind thousands of magnitudes of visual extinction and is only observable at far-infrared to radio wavelengths, providing indirect information about the stars themselves.
As soon as massive stars begin to emit UV photons and start to develop a stellar wind, the surrounding molecular cloud evaporates and the stars become visible at shorter wavelengths.

For the direct detection and the study of the recently formed massive stars, the near-infrared is the most suitable regime.  The emission of warm dust, the dominant source of emission at mid-infrared wavelengths, is much fainter in the near-infrared and the actual photospheres of the high-mass stars can be probed.  The success of this approach is demonstrated by several near-infrared imaging studies \citep[e.g.][Kaper et al. in prep.]{Hanson97,Blum00, Feldt03, Alvarez04,BikThesis}.  These imaging studies reveal the stellar content hidden in those star-forming regions, but also show the limitations, as  these studies are limited to regions with A$_V$ less than $\sim$ 30 mag. 
Near-infrared spectroscopy,  however, results in a much less ambiguous identification and classification of the massive star content \citep[e.g.][]{WatsonSpectra97, Blum00, Hanson02, Kendall03, Ostarspec05, Puga06}. These studies demonstrate the diagnostic power of near-infrared spectroscopy in characterizing the young, massive, stellar population.

Previous near-IR spectroscopic studies only considered a limited number of stars per cluster.  This incomplete spectroscopic sampling of the stellar population,  as well as the complexity of the fields, prevents any systematic investigation of the stellar content in the context of their environment.  A systematic study on how the stellar content depends on the cluster properties like total luminosity,  evolutionary phase as well as their environment requires a full spectroscopic census of the high-mass stellar content. 

We started a program to obtain a full near-infrared spectroscopic census of the stellar content of a sample of high-mass star-forming regions by means of integral field spectroscopy using SINFONI at the VLT. Integral field spectroscopy provides a spectrum of every spatial element in the field of view. This yields  not only  a spectrum of all the stars in the observed area, but also emission line images  of \HII\ regions, photon dominated regions (PDRs), and outflows. 
Supporting \emph{Spitzer} IRAC and MIPS imaging reveal the more deeply embedded objects and enable the classification of young stellar objects via their colors.

This paper is the first in a series of studies presenting our integral field data set of 10 high-mass star-forming regions. As subject of this first study we chose the region RCW34, associated with the Vela Molecular Ridge. The region consists of a well developed \HII\ region interacting with a dense molecular cloud to the North. 

The interaction between the \HII\ region and the molecular cloud suggests that a triggered star formation scenario could be  applicable to RCW 34.   Three different mechanisms are discussed in the literature, all resulting from the presence of an expanding \HII\ region inside a molecular cloud \citep{Elmegreen98,Deharveng05}. The first mechanism relies on the presence of pre-existing molecular clumps. The high pressure of the \HII\ region can result in the collapse of those clumps and the creation of cometary globules  forming a bright rim. The second mechanism also produces such globules and bright rims, however the clumps are not pre-existing but created by sweeping up gas by the \HII\ region. The cores become unstable on short time scales and are slowly eroded away by the \HII\ region.
 The third  mechanism is the "collect and collapse" mechanism \citep{Elmegreen77}. Similar to the second mechanism, the \HII\ region sweeps up neutral gas; however, in this scenario the neutral layer becomes unstable on longer timescales and very massive cores are formed. When these massive cores collapse they form new star clusters.
 
We will determine the properties (spectral-type, mass, age) of the high- and intermediate mass stellar population of RCW34 by means of near-infrared integral field spectroscopy and \emph{Spitzer}/IRAC images.  The cluster environment is characterized via \HII, \hei, and H$_{2}$  emission lines. The three triggering mechanisms are tested on the observed properties of RCW 34 and it turns out that an unusual mode of triggering might be at work here.

\subsection{RCW 34}

The \HII\ region RCW 34 \citep{Rodgers60} (IRAS 08546-4254  or Gum 29 \citep{Gum55})  is located in the Vela Molecular Ridge (VMR)  cloud  C \citep{Liseau92}. The region is an optically visible \HII\ region of about 2\arcmin $\times$ 2\arcmin\ in extent and slightly offset from the main  dense cores defining the VMR cloud C detected by CO observations \citep{Murphy91,Yamaguchi99}.  Due to a similar velocity (v$_{\mbox{lsr}}$ =  6 \kms), \citet{Murphy91} argue that RCW34 is likely part of the C-cloud. A similar velocity has been measured from the CS line   \citep[v$_{\mbox{lsr}}$ = 5.5  \kms,][]{Bronfman96}, confirming the association with cloud C, which would place the region at a distance somewhere between 1 and 2 kpc.

Herbst (1975) and   \citet{vandenBergh75}  allowed that the exciting star of RCW 34 might 
  be more distant, but favored an association with the Vela-R2 
  R-association at a distance of 870 pc. 
     Optical spectroscopy by 
  Heydari-Malayeri (1988) implies  that the central source   vdBH 25a \citep{vandenBergh75} is of spectral type 
 O8.5V.  This 
  suggests a distance of 2.9 kpc, supporting the idea RCW 34 is 
  more distant than Vela-R2.  In \S 4.1 we use near-infrared 
  spectroscopy of the cluster OB stars to derive a distance
  in rough agreement with Heydari-Malayeri (1988).  Heydari-Malayeri
  also detected a North-South velocity gradient in the ionized gas,
  which he suggested to be evidence that RCW 34 is an HII region 
  undergoing a champagne flow.  This is supported by Deharveng et al. 
  (2005) who conclude, based on MSX and near-infrared observations,
  that the region is not the result of the collect and collapse 
  triggered star formation scenario.

Several signs of active star formation are present in RCW 34. Two water masers have been reported \citep{Braz83,Caswell89} as well as a  class II methanol maser  \citep{Chan96,Walsh97}. Associated with the water masers, CS emission has been found  \citep{Zinchenko95} with an estimated mass of 1320 M$_{\sun}$. \citet{Gyulbudaghian07} reported the detection of  1.2 mm continuum emission towards RCW34.

The outline of the paper is as follows;  Section 2 outlines the data reduction of the SINFONI and \emph{Spitzer}/IRAC data sets used in this paper. In section 3, the interaction between the \HII\ region and the cluster surroundings is studied using the IRAC data as well as the SINFONI emission line maps. Section 4 focuses on the stellar spectra obtained in the SINFONI field; spectral types, luminosity class as well as extinctions are derived, and in Section 5 the stellar content is discussed and an age estimate is obtained for the NIR cluster in RCW 34. In Section 6 we discuss the  complete region and combine all the different properties to derive a possible formation scenario for RCW 34, and Section 7 summarizes the main conclusions of the paper.

\section{Observations and data reduction}

\subsection{SINFONI observations}
Near-infrared $H$- and $K$-band observations were performed using the Integral Field instrument SINFONI \citep{Eisenhauer03,Bonnet04} on UT4 (Yepun) of the VLT at ESO, Paranal, Chile. The observations of RCW 34 were performed in service mode  between February 4 and 9, 2007.  The non-AO mode of SINFONI, used in combination with the setting providing the widest field of view (8\arcsec $\times$8\arcsec), delivers a spatial resolution of 0.25\arcsec\ per slitlet.  The typical seeing during the observations was 0.7\arcsec\ in $K$. The H+K grating was used, covering these bands with a resolution of R=1500 in a single exposure.

We observed RCW 34 with a detector  integration time (DIT) of 30\,s per pointing to guarantee a good  S/N ($\sim$ 70) for the early B-type stars within this cluster. RCW 34 is much larger than the field of view of SINFONI, the near-infrared extent of the \HII\ region is 104\arcsec $\times$ 60\arcsec\ on the sky.  The observations were centered on the central O8.5 V star (vdBH 25a) at coordinates: $\alpha$(2000)= 08$^{h}$56$^{m}$ 28$^{s}$.1, $\delta$(2000) = -43$^{\circ}$05\arcmin55.6\arcsec. We observed this area with SINFONI using a raster pattern, covering every location in the cluster at least twice.
The offset in the   East-West direction was 4\arcsec\ (i.e.\ half the FOV of the detector) and  the offset in the North-South direction was 6.75\arcsec. 
A sky frame  was taken every 3 minutes using the same DIT as the science observations. The sky positions were chosen based on existing NIR imaging in order to avoid contamination. Immediately after every science observation, a telluric standard of B spectral type was observed, matching as close as possible the airmass  of the object. This star is used for the removal of the telluric absorption lines as well as for flux calibration.  The stars chosen as telluric standards were HIP 048652, HIP 045117, HIP 052202 and HIP 040629, one for each of the observing blocks.

\subsubsection{Data reduction}


The SINFONI data were reduced using the SPRED (version 1.37) software developed by the MPE SINFONI consortium \citep{Schreiber04,Abuter06}. 
This software was used to apply all the basic steps of the data reduction like flat-field correction, dark removal, distortion correction, wavelength calibration and correction of bad pixels.  Special care was taken to remove bad pixels from the calibration frames. 
The construction of the final calibration data was done in three iterations by first constructing a preliminary calibration set used to detect and remove the bad pixels using the 3D information. The cleaned raw files are used to construct a new calibration set and the procedure is repeated.

The bad pixels on the science data are removed using the same 3D interpolation method. However, due to severe cross talk between the pixels, the surrounding pixels are affected at lower intensity  and are not properly removed by the algorithm. Therefore  we apply an additional bad pixel removal later in the reduction process.  The cleaned science and sky frames are reconstructed into 3D cubes using the calibration data. Also the dark frames are subtracted  and the cubes are corrected for atmospheric diffraction. We applied the procedure described by \citet{Davies07} to remove OH line residuals.

In the reduced cubes we discovered a drift of the stellar position versus wavelength. The shift is strongest in $K$ (up to 2 pixels) and resulted in an elongation of the  stars in the collapsed image. We measured the position of the telluric standard star versus wavelength using a 2D gaussian and fitted a spline to the change of position in both spatial directions. This spline fit was used to correct the science cubes. The residual measured on some bright stars in the science frames was within 0.5 pixels, which is acceptable for our purposes.

The last step in the reduction process is flux calibration and removal of the telluric absorption lines. This is done using the extracted spectrum of a B  type standard star,  using an aperture radius of 6 pixels (0.75\arcsec). After the extraction, the correction for the telluric absorption lines in the science spectra  is done in two steps: First, the telluric standard star is corrected using a high signal-to-noise atmospheric template spectrum \footnote{taken from the SINFONI website and obtained at NSO/Kitt Peak Observatory}.  This high S/N spectrum is not taken at the same airmass and atmospheric conditions as our science spectra, but provides a first rudimentary correction  in  order to reduce the absorption contamination of the telluric standard star.

Second, the
Brackett and \hei\ lines are fitted with a Lorentzian profile. The line-fits are subtracted
from the original telluric  standard star spectrum, leaving an atmosphere
transmission spectrum.
The flux calibration of the spectra uses the 2MASS \citep{Cohen03} magnitude of the standard stars.
As we only use normalized spectra for the point sources and flux ratios for the emission line spectra, the uncertainty in the flux calibration due to atmospheric variations is  not critical.

The correction of the atmospheric absorption lines and the flux calibration are performed on the 3D cubes; the spectrum of every spatial pixel is divided by the spectrum of the telluric standard star and flux calibrated.  Finally, the individual 3D cubes of one observing block (between 50 and 60 cubes) were combined into one large 3D cube. For RCW 34, four of those large cubes are needed to cover the near IR emission of the cluster.

\begin{figure*}
\plotone{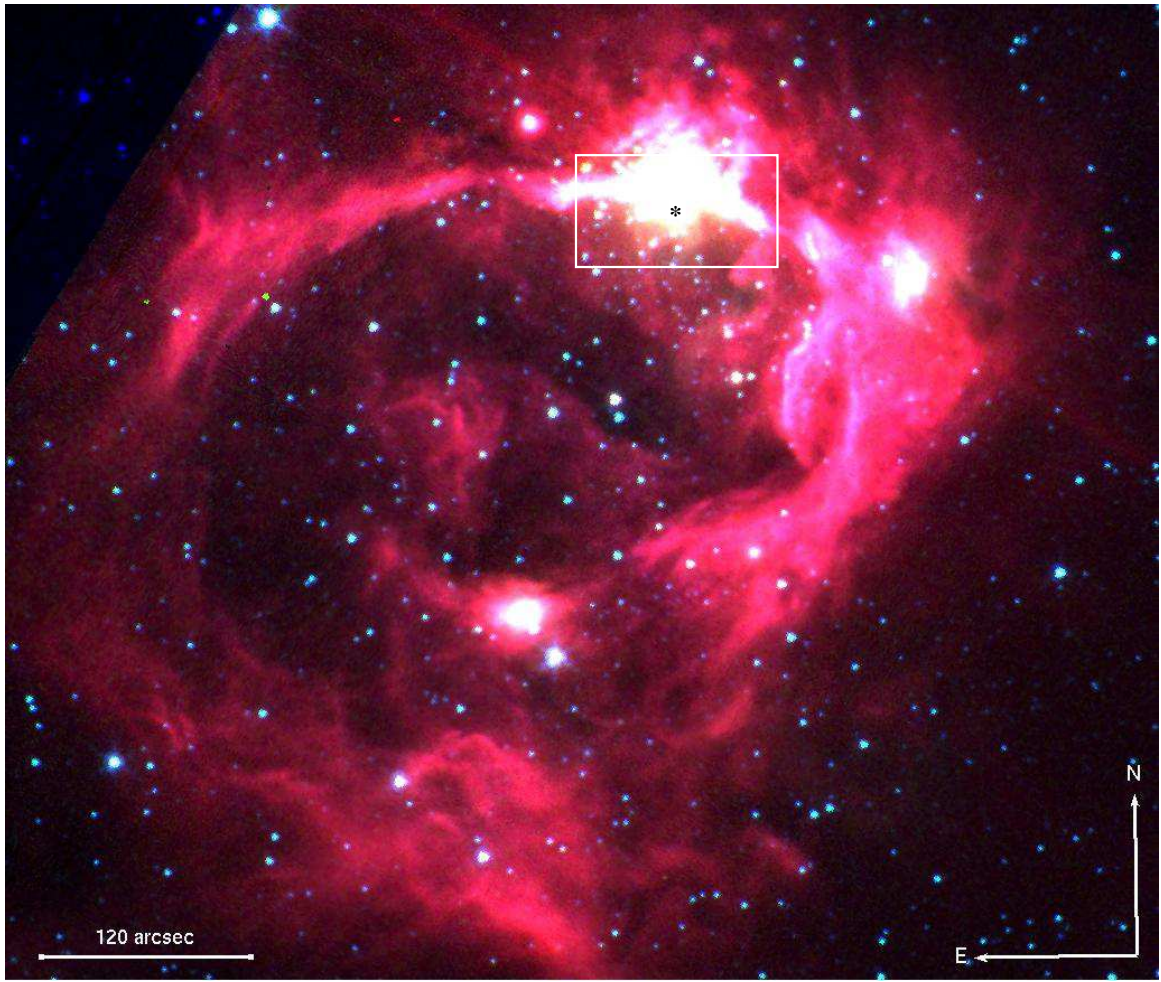}
\caption{\emph{Spitzer}/IRAC 3 color image of RCW 34 and its surroundings (blue: 3.6 \micron, green: 4.5 \micron, red: 8 \micron, color version online).  The top left corner is only covered at 3.6 \micron.   The area delimited by the white box is the area covered by the SINFONI observations. The location of the central star  of RCW 34, vdBH 25a ($\alpha$(2000)= 08$^{h}$56$^{m}$ 28$^{s}$.1, $\delta$(2000) = -43$^{\circ}$05\arcmin55.6\arcsec.), is shown with an asterisk. The \HII\ region surrounds vdBH 25a and is fully covered by our SINFONI observations.  \label{fig:spitzer_ima}} 
\end{figure*}

Before extracting the  point source spectra from the science cubes, an attempt was made to remove the extended background emission from the data. This background emission is primarily nebular emission from the HII region and the PDR (Brackett lines, \hei, H$_{2}$). A median filter with a kernel of 5 pixels is used to remove the extended nebular component in the cubes. We applied this median filter 3 times to remove any high frequency variations present which could be due to stars. This median filtering has only a small effect on the stellar continuum;  the contribution in the total flux is less than 1 \%.  The point sources were extracted using a circular aperture with a radius of 6 pixels, analogous to that of the standard star.

Several maps centered on emission lines as well as continuum maps are created from the combined SINFONI data cubes.
The maps are made by summing 4 to 7 slices of the data cube in the wavelength direction. The images in the spectral lines are corrected for continuum emission from the stars by interpolating each pixel of the adjacent continuum images (bluewards and redwards of the line, respectively). This results in a clean subtraction  for most of the stars. However, for the brightest stars in the maps, residuals remain present. An interpolation of  the line emission a bit further away from the stars successfully restored  these residuals.

We  created velocity maps of the strongest emission lines, \brg\ and H$_{2}$, for every (spatial) pixel in the
data cube. The central velocity of the \brg\ and   H$_{2}$ emission
line is derived from a gaussian fit to the line profile.  To obtain better accuracy in the wavelength calibration we performed the same fitting procedure to a nearby OH skyline. Finally, the velocities were converted to the local standard of rest. The errors were calculated by quadratically summing the fitting errors on the nebular  line and the OH line and are on the order of 10 - 15 \kms.

\subsection{IRAC data reduction and photometry}

Additional data obtained with IRAC \citep{Fazio04} on board of the \emph{Spitzer} satellite have been retrieved from the \emph{Spitzer} archive (PID: 40791, PI: Majewski). The data were obtained on January 31, 2008 in a mapping mode, covering a large part of the outer galaxy. We selected the frames covering the area around the cluster as well as an adjacent  control field.
Each frame has an exposure time of 1.2 seconds and every position is covered twice. The raw frames were processed using the standard pipeline version S17.02 to produce the basic calibration data set. This data set was processed further using MOPEX version 16.2.1 \citep{Makovoz06}. 

As the IRAC data are undersampled, aperture photometry is preferred above psf-fitting photometry. Source detection and photometry were performed using the Daophot package under IRAF.  Point sources with a peak level of more than 5 times the RMS of the background were detected. Aperture photometry was applied with an aperture of 3 pixels (3.7\arcsec) and a sky annulus between 3 (3.7\arcsec) and 7 (8.5\arcsec) pixels. These small values were chosen due to the  variable extended emission. The appropriate aperture corrections and zeropoints were adopted (Tables 5.1 and 5.7 of the IRAC Data Handbook) to obtain the magnitudes of the objects.

\section{Cluster surroundings}

  NIR imaging (Bik 2004; Deharveng et al. 2005) shows an \HII\ region
 extending about $1.5' \times 1'$ around the central star vdBH 25a.
 The IRAC images (Fig. 1) show a much larger star-forming region than
 is visible in the NIR.  The region extends to the South and East and
 shows an open bubble structure.  The total size of the mid-IR region
 is about $6' \times 9'$.  This structure looks similar to the (partial)
 bubbles identified in the GLIMPSE survey (Churchwell et al. 2006, 2007).
 
In this section we will determine the nature of the stars located inside the bubble using the IRAC photometry and investigate whether the objects in the bubble are related to those detected in the \HII\ region. We analyze the extended emission seen in the SINFONI observations and derive properties of the \HII\ region and the surrounding molecular cloud.

\begin{figure}
\plotone{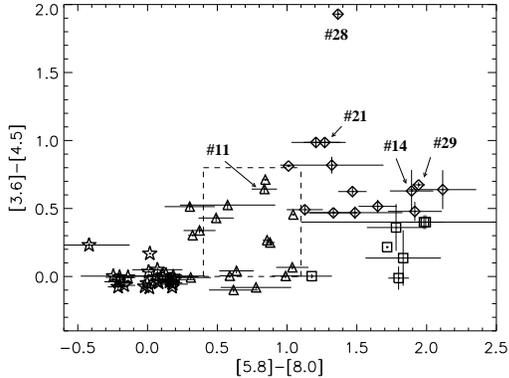}
\caption{ \emph{Spitzer}/IRAC color-color diagram. The star-shaped points represent stars without infrared excess, class II stars are plotted as triangles inside the squared box \citep{Megeath04}. The diamond symbols represent the location of the protostars (class 0/I) and the squares show the location of the class I/II sources.  The numbered objects are located inside the SINFONI field and are discussed in more detail in Sect. 4.3. \label{fig:spitzer_ccdiag} }
\end{figure}

\begin{figure*}
\epsscale{0.85}
\plotone{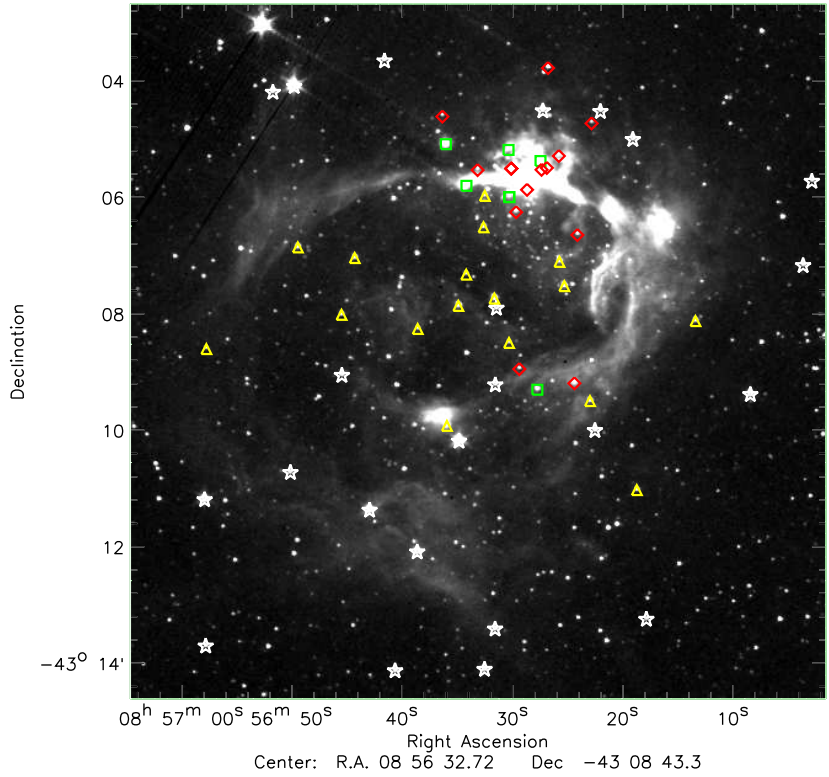}
\caption{ \emph{Spitzer}/IRAC 3.6 \micron\ image of RCW 34 and its surroundings. The different classes of objects as derived from the IRAC color-color diagram  are overplotted. The symbols used are identical to those in Fig. \ref{fig:spitzer_ccdiag}. \label{fig:spitzer_objects}} 
\end{figure*}

\subsection{Spitzer/IRAC imaging of the cluster and its surroundings}

A 3-color \emph{Spitzer}/IRAC image of the region is shown in Fig. \ref{fig:spitzer_ima} and Fig. \ref{fig:spitzer_ccdiag} shows a [3.6]-[4.5] vs. [5.8]-[8.0] color-color diagram of the stars in the field. Objects not detected in all bands are not included in the analysis. We have not plotted the stars in the control field  which only contains stars without IR excess. With the color criteria proposed by \citet{Megeath04} and \citet{Allen04}  we can identify several classes of objects. Stars without IR-excess have colors close to 0.0.  The IRAC colors are insensitive to the effective temperature of the photospheres.  These objects, plotted as star-shaped symbols,  could be fore- or background late-type stars as well as earlier-type main-sequence stars associated with the star-forming region. 
 The objects with [3.6]-[4.5] between 0.0 and 0.8 and [5.8]-[8.0] between 0.4 and 1.1 are classified as class II objects. This region is marked with a box, and the objects showing class II colors are plotted as triangles. Objects with [3.6]-[4.5] $>$ 0.8 and [5.8]-[8.0] $>$ 0.0 or [3.6]-[4.5] $>$ 0.4 and [5.8]-[8.0] $>$ 1.1 are classified as protostars (class 0 and I objects, diamonds), objects with [5.8]-[8.0] $>$ 1.1 and [3.6]-[4.5] between 0.0 and 0.4 are  classified as class I/II sources (square symbols).  The 5 numbered objects are located in the SINFONI field and are discussed in more detail in Sect. 4.3. and 5.3.

The class I/II identification should be considered with caution. \citet{Megeath04} show that these objects are likely class II sources with the 8 \micron\ flux contaminated by variable extended emission. Fig. \ref{fig:spitzer_ima} shows that these objects  are located on top of a strong diffuse background.  Therefore, we consider them as class II objects. Additional sources of contamination, especially for the detection of protostars, could be planetary nebula, AGB stars or background galaxies \citep{Whitney03}. However, the absence of sources with these colors in the control field suggests that the contamination is negligible.

We have plotted the different classes on the IRAC 3.6 \micron\ image (Fig. \ref{fig:spitzer_objects}) to study their potential relation with the extended MIR emission of the bubble.  The locations of the stars without infrared excess (white, star-shapes) are not related to the extended emission, but evenly distributed over the field outside the bubble.  
 These objects are most likely field stars.

The location of the YSOs shows a clear correlation with the bubble and the extended emission. Most of the class II sources (yellow triangles) are located inside the bubble. The protostars (red triangles) are spatially associated with the extended MIR emission. Most of these objects are located near the bright arc in the north, but also some objects are associated with the fainter extended emission in the southern part of the bubble.  Although no obvious cluster is visible inside the bubble, the class II  objects located inside the bubble suggest a recent star formation event.  This might have been responsible for the creation of  the bubble \citep{Churchwell06}.  The location of the younger protostars in the outer regions of the bubble indicates the presence of  a more recent generation of stars  near the edge of the bubble.

\begin{figure*}
\plotone{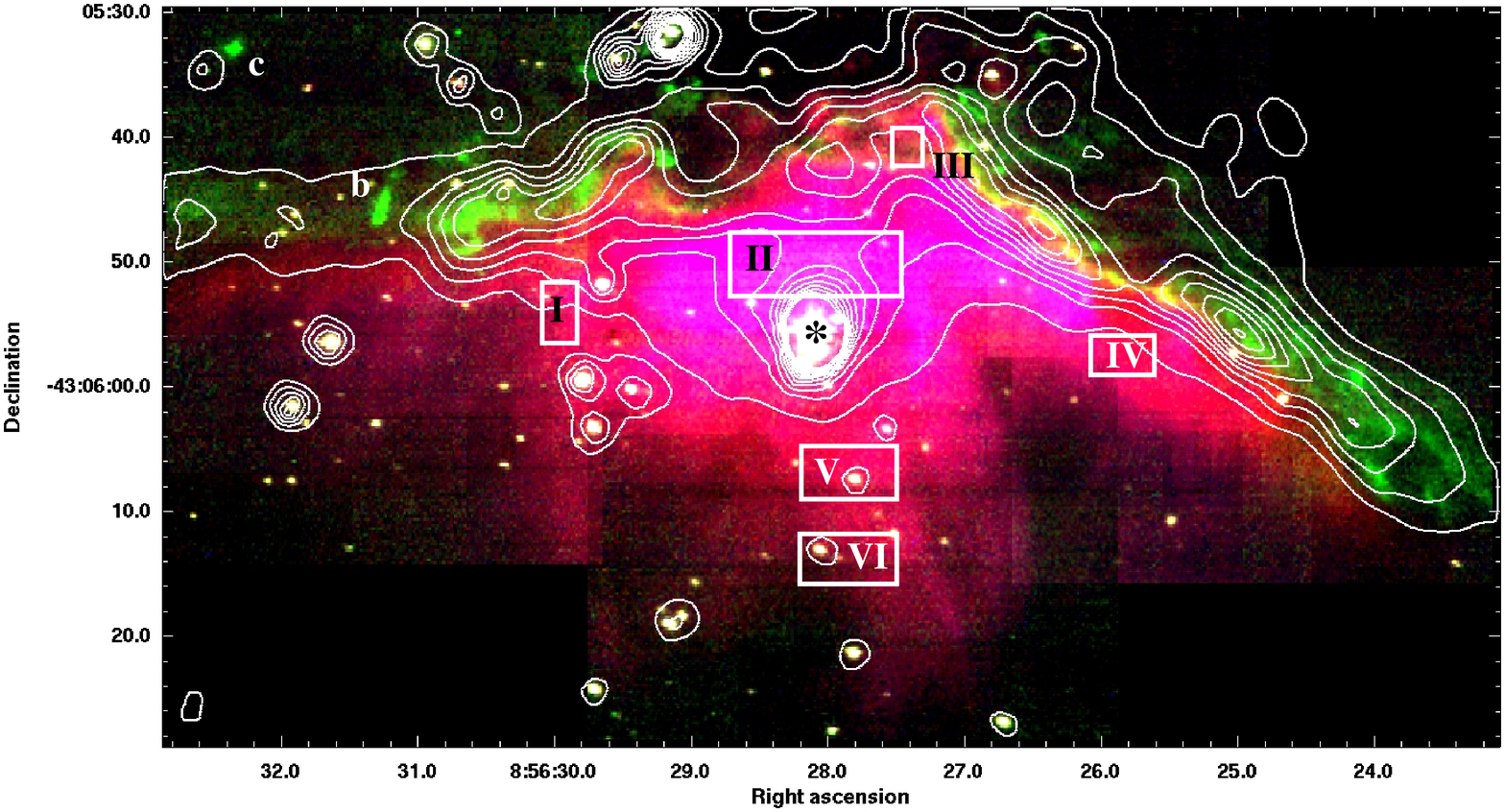}
\caption{Three-color image created from the SINFONI line maps of RCW 34 (color version online). Red: \brg\ (line + continuum) emission, green: H$_{2}$ (2.12 \micron, line  + continuum) and blue: \hei\ (2.058 \micron, line + continuum).  The horizontal striping is a data reduction effect.  Overplotted in contours is the 3.6 \micron\ image. The letters b, c mark the parts of the H$_{2}$ emission (region (a) is not annotated but comprises the entire arc North of the \HII\ region) for which excitation diagrams are constructed to identify the nature of the emission (Fig. \ref{fig:H2_diag}). The six boxes show the regions from which  spectra are  extracted to derive the extinction and the \hei /\brg\ ratio. The asterisk shows the location of the main ionizing source vdBH 25a. 
\label{fig:lineimage}}
\end{figure*}

\subsection{SINFONI line and continuum  maps}

The SINFONI observations are centered on the ionizing source, vdBH 25a,  and cover both the \HII\ region as well as the bright arc seen in the IRAC images.  Fig. \ref{fig:lineimage}  shows the 3 color composite of 3 linemaps along with contours of the IRAC 3.6 \micron\ image. This image reveals the stellar content of this region  and the overlay  shows that several near-infrared sources are  also detected at IRAC wavelengths. The region is dominated by the two bright stars in the center of the field (one of them is vdBH 25a at the location asterisk symbol) and a relatively sparse cluster of about 130 stars down to $K$=17 mag, surrounding the central region. For all the point sources visible in this image a SINFONI $H$- and $K$-band spectrum is available and a spectral classification is obtained (Sect. 4).  The \HII\ region is traced by \brg\ (2.16 \micron, red) and \hei\ (2.058 \micron, blue) emission and the surrounding PDR by the molecular H$_2$ emission (2.12 \micron, green). Just North of the position of the central star, the \brg\ emission is strongest.  Even further to the north,  the \brg\ and \hei\ emission drop abruptly and the H$_2$ emission begins, indicating the interface
between the ionized and molecular gas, coincident with the bright arc
seen in the IRAC images.


\subsubsection{HII region}

We have extracted spectra at  six locations covering most of the \HII\ region (annotated with numbers  I - VI in in Fig. \ref{fig:lineimage}).  Table  \ref{tab:HII}  gives the position and size of the six extracted regions.

\begin{figure*}
\plotone{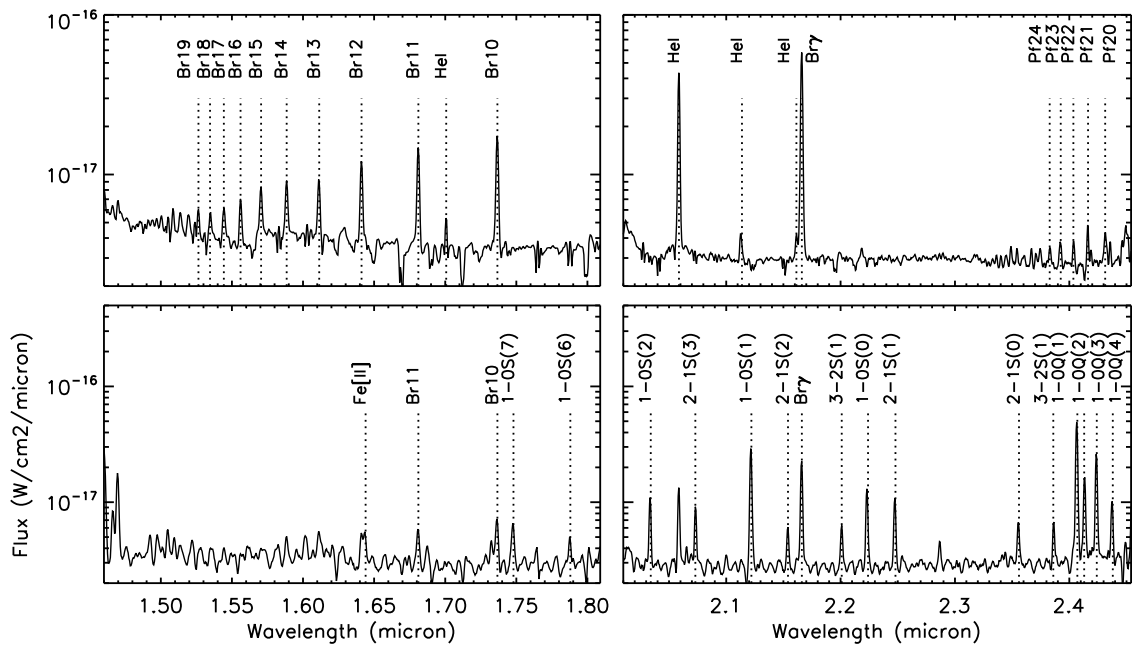}
\caption{\emph{Top panels:} $H$- and $K$-band SINFONI  spectra of the \HII\ region in RCW 34. The spectra are drawn from a region just North of the ionizing stars  (region II in Fig. \ref{fig:lineimage}).  The spectra show numerous hydrogen lines of which \brg\ (2.166 \micron) is the most prominent. Also several \hei\ lines are visible, suggesting that the central star is of spectral type late O. \emph{Bottom panels:} Spectra of the photo dominated region (PDR), North of the \HII\ region (region (a)). The spectrum is dominated by H$_{2}$ lines.  Also some faint [Fe II] emission is present.  This line can be formed in J-shocks and also in the low density material
between the \HII\ region and the PDR.
 \label{fig:brg_spectrum}}
\end{figure*}

As an example, the upper panels of Fig. \ref{fig:brg_spectrum} show the spectra extracted from  region  II, just North of the ionizing star.

To check the hypothesis of a champagne flow, we estimate the column density of the molecular cloud in which the \HII\ region is embedded. In an \HII\ region, the relative line intensities of the hydrogen recombination lines are mostly determined by the Einstein coefficients, with a very weak dependence on density and electron temperature \citep{Storey95}. For example, the ratio of the Br10 (1.73 \micron) and the \brg\ (2.16 \micron) lines is predicted to be 0.33.  Deviations from this value are caused by foreground extinction.  We have converted the measured Br10/\brg\ ratio to A$_{V}$ using the \citet{Cardelli89} extinction law.
The derived A$_{V}$ values for the six extracted regions are listed in column 7 of Table \ref{tab:HII}. The bulk of the \HII\ region has an A$_{V}$ of around 5-6 mag   (regions I, II and IV).  In region III, on the border with the molecular cloud,  we measure an A$_{V}$ of 15.9 mag, while towards the South the extinction declines to 4.4 mag and 1.5 mag for regions V and VI respectively.  This extinction gradient  can be explained as a density gradient in the surrounding molecular cloud and is consistent with the presence  of a champagne flow.

\begin{deluxetable*}{lllrlrr}
\tablewidth{0pt}
\tablecaption{Properties of the \HII\ gas \label{tab:HII}}
\tablehead{
\colhead{Region} &
\colhead{R.A. (2000)} & 
\colhead{DEC. (2000)} &
\colhead{size } & 
\colhead{$v_{lsr}$} & 
\colhead{\hei /\brg} & 
\colhead{A$_{V}$} \\
\colhead{} &
\colhead{(h m s)} & 
\colhead{($^\circ$\ \arcmin\  \arcsec)} &
\colhead{(\arcsec)} & 
\colhead{(\kms)} & 
\colhead{} & 
\colhead{(mag)} 
}
\startdata
I	&08 56 29.8&-43 05 52.4&2.4     $\times$  4.6& 18 $\pm$ 13	& $<$ 0.13                    & 5.3 $\pm$ 0.2	\\
II	&08 56 28.1&-43 05 50.8&13.7  $\times$       4.9& 16 $\pm$ 8	&  0.038$\pm$	0.002 & 5.1 $\pm$ 0.1	 \\
III	&08 56 27.3&-43 05 39.2& 2.3    $\times$     2.7& 16 $\pm$  9	& $<$ 0.17                   & 15.9 $\pm$ 0.6 \\  
IV	& 08 56 25.7 & -43 05 55.7& 5.0    $\times$     3.0& 10 $\pm$12	& 0.032 $\pm$ 0.006  & 5.6 $\pm$ 0.2	\\
V	&08 56 27.7& -43 06  05.0&7.7    $\times$     4.0& 13 $\pm$ 9	& 0.047 $\pm$ 0.004  & 4.4 $\pm $ 0.1	\\ 
VI	&08 56 27.7&-43 06 11.9& 7.7     $\times$    3.7& 11 $\pm$ 10  & 0.06    $ \pm$ 0.01	  & 1.4 $\pm $ 0.2 
\enddata
\end{deluxetable*}

 The (extinction corrected) ratio of the 1.70 \micron\  \hei\ line over  \brg\ is dependent on the temperature of the ionizing star and can be used to constrain its spectral type \citep{Lumsden03}. The extinction correction line ratios for the six regions are given in column 6 of Table \ref{tab:HII}. 
 The maximum ratio is measured just South of the location of the ionizing stars, and has a value of 0.047 $\pm$  0.004 in region V and 0.06 $\pm$ 0.01 in region VI.  Comparing this value to the predictions of  \citet{Lumsden03}, an effective temperature for the central star around 35000\,K is expected, which corresponds to a spectral type between O8\,V and O9\,V \citep{Martins02}, consistent with the optical spectral type determination of \citet{Heydari88}.

\subsubsection{PDR and outflows}

The IRAC detections of class 0 and I sources show that stars are still
forming close to the bright arc. An additional argument for  active star formation is the presence  of outflows, detectable by  H$_{2}$ line emission (bottom panels Fig. \ref{fig:brg_spectrum}). 

The H$_{2}$ emission in the near-IR can be either  thermally or non-thermally excited. The thermal emission originates from shocked neutral gas being heated up to a few 1000 K (outflows), while the non-thermal emission is due to fluorescence by non-ionizing  UV photons between 912 and 1108 \AA. These two mechanisms  can be distinguished as they preferentially populate different energy levels giving rise to different line ratios. The fluorescence mechanism preferably excites the higher vibrational levels, while for thermal excitation the lower levels are preferred \citep{Davis03}.

\begin{figure*}
\plotone{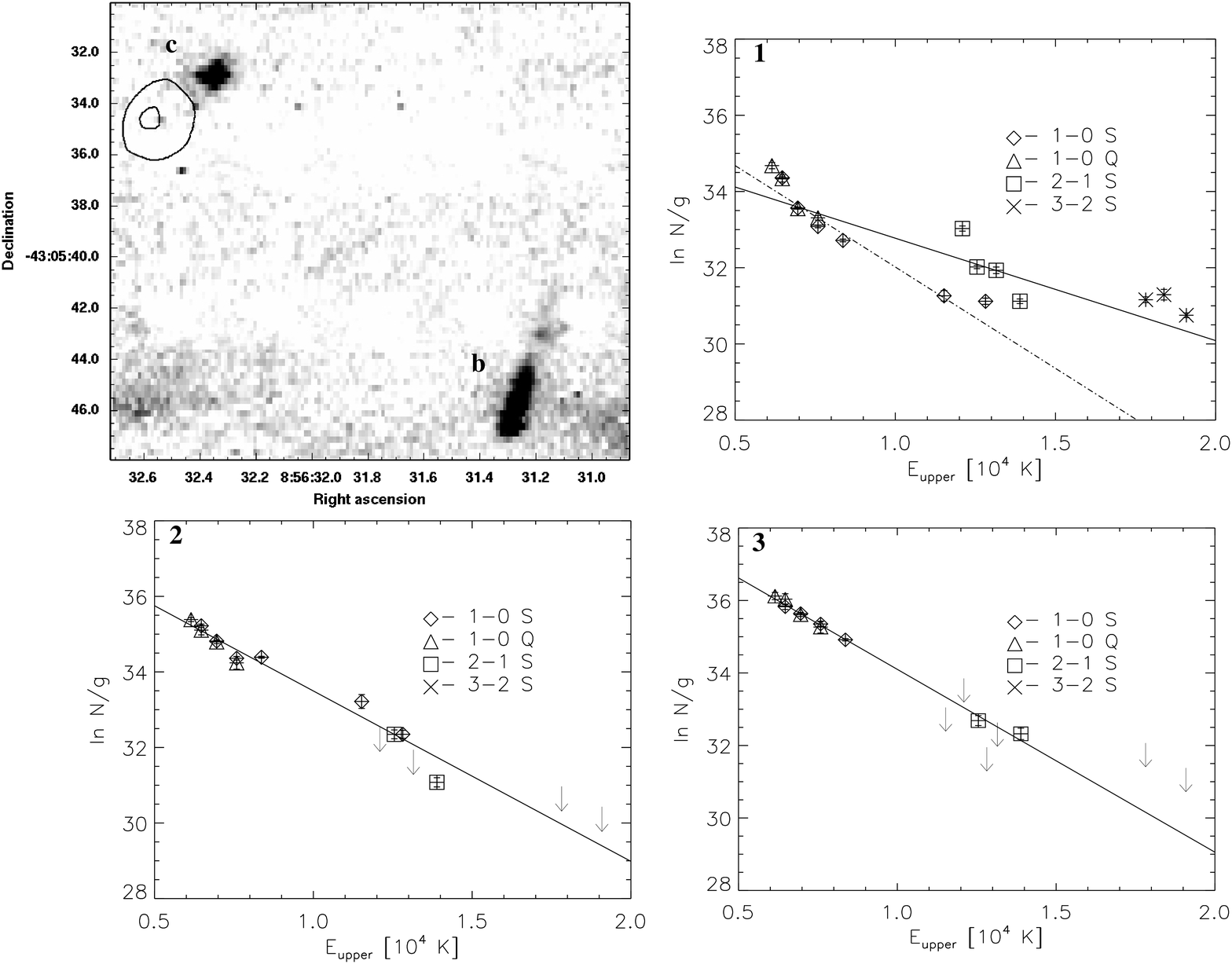}
\caption{Top left: zoom on the SINFONI H$_{2}$ map highlighting the two candidate outflow regions (b) and (c). The contour is the 3.6 \micron\ image highlighting source \#28, the possible driving source for region (c).  Panels 1 (region (a)), 2 (region (b)) and 3  (region (c))show the excitation diagrams of 3 different regions in the PDR identified in Fig \ref{fig:lineimage}.  The solid line in all the diagrams is a single temperature fit to all the lines detected, while for the dash-dotted line in panel a only the 1-0 S lines are included in the fit.  The excitation diagram of region (a) is that of the complete arc North of the \HII\ region. The excitation diagram suggests that UV fluorescence is the excitation mechanism in this area.  Regions (b) and (c) are  most likely outflows as their excitation diagrams suggest thermal excitation. \label{fig:H2_diag}}
\end{figure*}

Fig. \ref{fig:lineimage} shows the location of the  2.12 \micron\  H$_{2}$ emission (green)  in relation to the emission from ionized gas (\brg\ and \hei). The H$_{2}$ emission has a very filamentary structure and  appears just North of the ionized gas, co-located with the bright arc in the IRAC images.  Additionally, two compact sources are visible in the north-east.  To identify the excitation mechanism of the arc of filamentary emission (region (a)) as well as that of the two compact regions ((b) and (c), top left panel of Fig. \ref{fig:H2_diag}) we construct excitation diagrams for these regions.

First, the line flux ratios are extinction corrected. The ratio of the 1-0 S(1) (2.12 \micron) over the 1-0 Q(3) (2.423 \micron) is insensitive to the ionization mechanism and has an expected value of 0.7 \citep{Turner77}. Therefore, the observed ratios are used for an extinction estimate in the same manner as discussed in Sect. 3.2.1 (Table \ref{tab:H$_{2}$}).  Region (a) covers a large area, 262 square arcsecs and shows a large extinction variation. The derived properties are dominated by the strongest emission regions, located close to the \HII\ region where the extinction is lowest. 

  Fig. \ref{fig:H2_diag} also show the excitation diagrams of the 3 regions.  In these diagrams, the measured column densities of the lines are plotted against the energy of the upper level (see \citet{Martin08} for a detailed description of this diagram). The total column density was calculated using the description of \citet{Zhang98}. 
Different vibrational levels are indicated with different symbols. Region  (a) has the most lines detected as it covers the largest area (the complete arc). For the spectra of regions (b) and (c), the weaker lines seen in the spectrum of region (a) are not detected and  3 $\sigma$ upper limits are given instead. The solid line in the diagrams is a single temperature fit to all the data points, while the dashed line in the diagram of region (a) represents a linear fit to only the 1-0 S lines.  For region (a) it is clear that the column densities are not represented by a single temperature gas, but that the line fluxes of the 2-1 and 3-2 vibrational levels are higher than expected from the 1-0 line fluxes. A likely excitation mechanism of the H$_{2}$ gas in the arc is fluorescence by non-ionizing UV photons. 
We have searched for spatial variations of the temperature and column density in region (a) by calculating the excitation diagrams for several sub-regions, but could not detect any significant deviations from the values of the total area.

The excitation diagrams of regions (b) and (c) are well-represented by
a single temperature.  We note, however, that the 3-2 lines are not detected 
in either region, and the 2-1 lines are mostly upper limits.  We conclude
that the excitation diagrams of both regions are consistent with shock
excited emission in an outflow.  Their elongated shapes, especially for
region (b), are also suggestive of outflow.  The linear fits of the
excitation diagrams also allow us to determine the column density and 
temperature of the emitting gas (Table  \ref{tab:H$_{2}$}).

\begin{deluxetable}{crrrr}
\tablecaption{Physical properties of the H$_{2}$ gas. \label{tab:H$_{2}$}}
\tablehead{
\colhead{Region} & 
\colhead{T$_{rot}$} & 
\colhead{N(H$_{2}$) } &
\colhead{A$_{V} $}&
\colhead{v$_{lsr}$}\\
\colhead{} & 
\colhead{ (K)} & 
\colhead{  (cm$^{-2}$)} &
\colhead{(mag)}&
\colhead{ (km/s)}
}
\startdata
a	&  3720 $\pm$  90&	3.6  $\pm$ 0.6 $\times$ 10$^{17}$& 9 &  +1 $\pm $ 14\\ 
b	&  2217 $\pm$	90 &	1.8  $\pm$ 0.4 $\times$ 10$^{18}$&  13 & -14 $\pm $ 14 \\
c	&  1980 $\pm$ 115 & 4.6  $\pm$ 2.0 $\times$ 10$^{18}$&   20 &-26 $\pm$ 16
\enddata
\end{deluxetable}

An additional constraint on the nature of the H$_{2}$ emission regions is the velocity of the different regions. 
The obtained velocities are listed in the last column of Table \ref{tab:H$_{2}$}. The two outflows have a clear blue shifted velocity compared to that of the extended PDR, with region c having the largest value of $-26 \pm 16$ \kms. These relatively low velocities as well as the absence  of  [Fe\,{\sc ii}] at 1.644 \micron\ suggest that both shocks are C-type shocks \citep{Stahler05}. The more destructive shocks (J-type) would have strong [Fe\,{\sc ii}] emission as well as larger shock velocities \citep{Hollenbach89}.  The velocity measured in the arc (region (a)) of $1 \pm 14$ \kms\  is more blue shifted than the velocity of the  \HII\ region, but  consistent with that of the Vela molecular cloud: 6 \kms.

For outflow (b) a driving source could not be identified in the IRAC images, most likely the source is deeply embedded and not visible at  8  \micron. For source (c), however,  we could identify a possible driving source. About 3\arcsec\ South-East of the H$_{2}$ emission a deeply embedded protostar is identified on the IRAC images with a very bright 4.5 \micron\ flux (source \# 28 in Fig. \ref{fig:spitzer_ccdiag}). The location of the source is shown in  Figs. \ref{fig:lineimage} and  \ref{fig:H2_diag} where the 3.6 \micron\ contours of source \# 28 are overlaid on the SINFONI images.

The excess 4.5 \micron\ emission is most likely coming from H$_{2}$ emission closer to the source, too deeply embedded to detect at near-infrared wavelengths.

\subsection{Summary}

Summarizing, RCW 34 consists of 3 distinct regions. A  large bubble is  detected in the IRAC images, with a cluster of class II stars inside.  At the northern edge of this bubble the \HII\ region is located, ionized by vdBH 25a (O8.5 V). The detected North-South extinction gradient as well as the H$\alpha$ velocity gradient found by \citet{Heydari88} suggest that the \HII\ region is a champagne flow.

 In between the molecular cloud and the \HII\ region a PDR is located traced by H$_2$ emission and 8 \micron\ emission. The star formation inside the molecular cloud seems to be more recent as masers, class 0/I sources and outflows are present in that area.

\section{SINFONI Near-Infrared spectroscopy}

\begin{figure}
\plotone{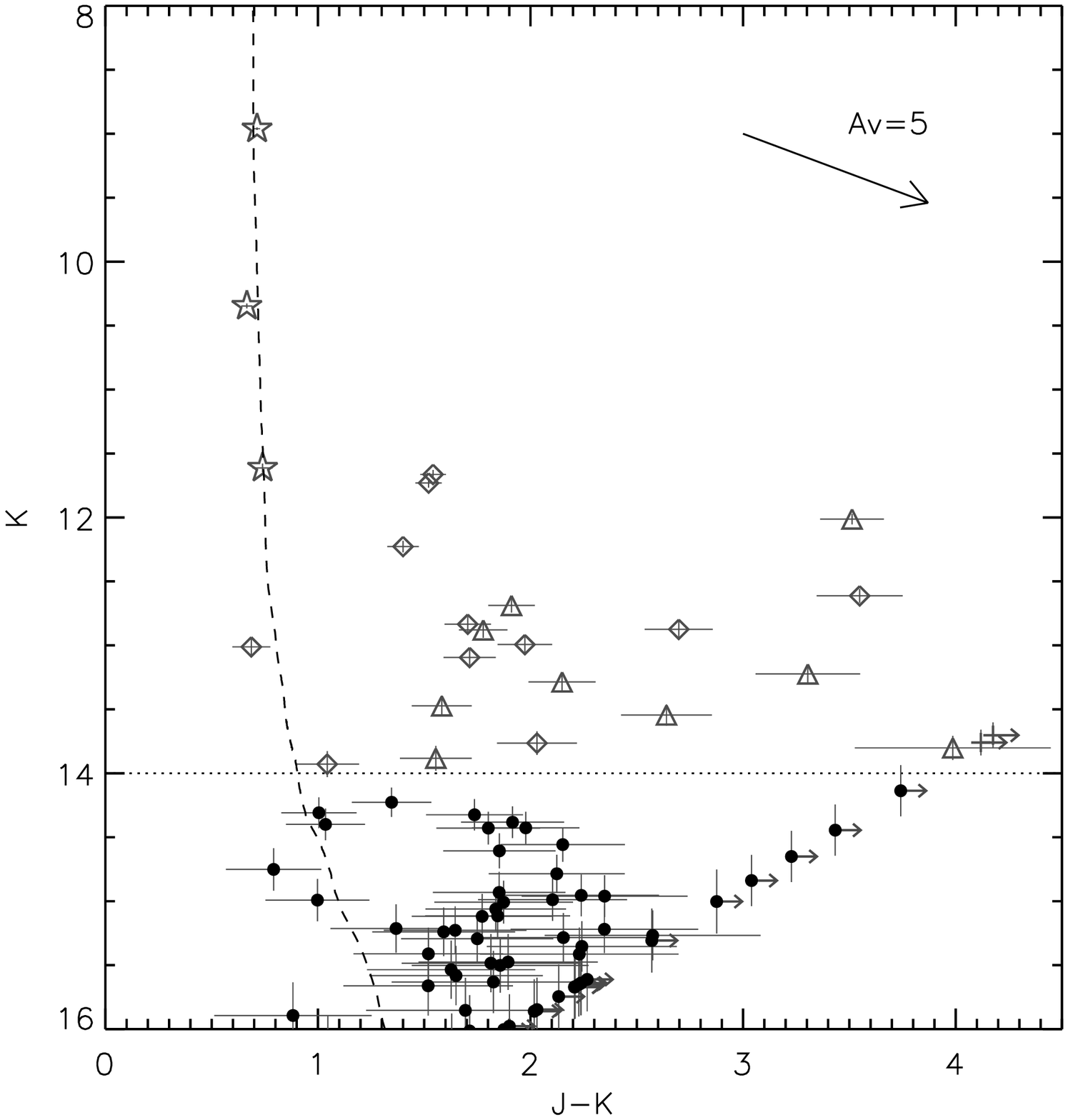}
\caption{Observed $K$ vs. ($J$-$K$)  color-magnitude diagram of the region in RCW 34 covered by SINFONI taken from \citet{BikThesis}. Stars brigher than $K$=14 mag  (horizontal dotted line) are marked with symbols representing their spectral types derived in Sect. 4.1 and 4.2 ( star symbols:  OB stars, diamonds: G stars, triangles:  K stars). The featureless stars are displayed as crosses. The dashed line represents the 2 Myr isochrone taken from \citet{LeJeune01}. \label{fig:cmd}}
\end{figure}

This section  focuses on the stellar content of the NIR cluster inside the \HII\ region RCW 34.  In total 134 stars have been identified between $K$=8.7 and 17 mag (see Fig. \ref{fig:lineimage}  for their location). Our SINFONI observations provide  an $H$- and $K$-band spectrum of each source with a spectral resolution of R=1500. Stars with $K$ $< $ 14 have high enough S/N spectra to obtain a reliable spectral type. This reduces the sample  to 26 stars, which will be discussed in the remainder of this paper.  

To estimate the stellar content, Fig. \ref{fig:cmd} shows the $K$ vs. ($J$-$K$)  color-magnitude diagram (CMD) of the area in RCW 34 covered with SINFONI taken from  \citet{BikThesis}, who imaged a 5\arcmin $\times$ 5\arcmin\ field using SOFI at the NTT telescope.  

The symbols in the plot represent the different spectral types derived later in this section. The CMD shows that the cluster consists of stars showing a large range in $J-K$ colors, which could be related to large extinction variations in the cluster.   
 In the remainder of the paper, the brightest star (vdBH 25a) is labeled \#1 and the label number increases with $K$-band magnitude. Table \ref{tab:master} provides the list of the 26 brightest stars, as well as some additional objects of special interest, and their properties derived in this paper.

\subsection{Early type stars}

\begin{figure*}
\plotone{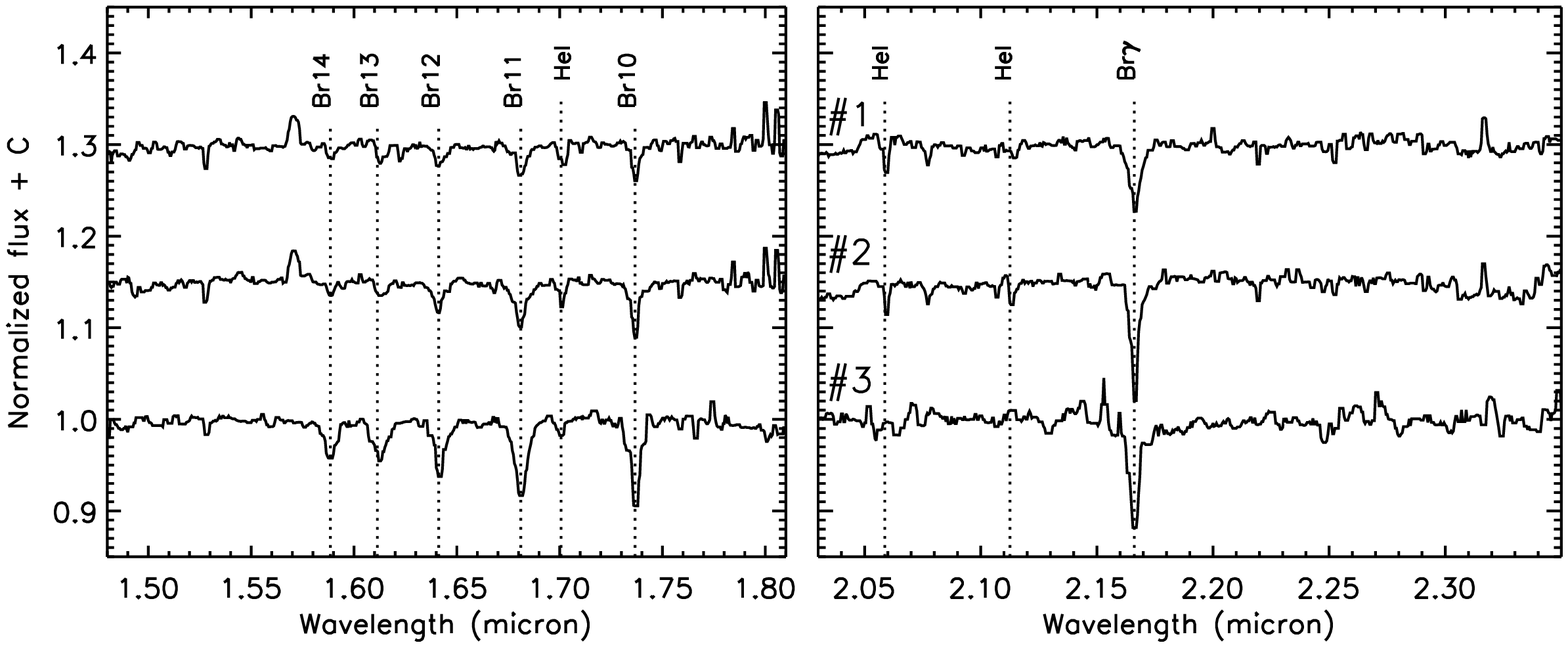}
\caption{Normalized SINFONI $H$-band (left panel) and $K$-band spectra of the three OB stars detected in RCW 34.  Their spectra are characterized by absorption lines of the hydrogen Brackett series  and \hei\ (dotted vertical lines). \label{fig:obstars}}
\end{figure*}

\begin{deluxetable*}{rrrrrrr}
\tablecaption{Equivalent widths and $K$-band spectral types of the OB stars in RCW 34.}
\tablehead{
\colhead{Star}      & 
\colhead{Br11 (1.68 \micron)}                          & 
\colhead{\hei\ (1.70 \micron)}                          &   
\colhead{\hei\ (2.11 \micron)}   &
\colhead{\brg\ (2.166 \micron)}            & 
\colhead{K-Spectral type} & 
\colhead{Optical Spectral type}
}
\startdata
1   &     2.05 $\pm$ 0.3 \AA                    & 0.91 $\pm$ 0.2  \AA                         &  0.4  $\pm$ 0.4 \AA                 & 4.43  $\pm$ 0.4 \AA                    & kO9-B1   & O8V - B1V       \\ 
2   &     2.90  $\pm$ 0.4    \AA                    & 0.75  $\pm$ 0.2 \AA                         &  0.6 $\pm$ 0.2  \AA                 & 6.07  $\pm$ 0.3 \AA                    & kB2-B3   & B1V - B2.5V    \\
3   &     5.79   $\pm$ 0.7   \AA                    & 0.76  $\pm$ 0.2 \AA                         &   $<$ 0.4                         & 6.06 $\pm$ 0.4  \AA                    & kB4-B7   & B3V - B7V 	 
\enddata
\label{tab:classification}
\end{deluxetable*}

The SINFONI H+K spectra of the three brightest cluster members (sources \#1, \#2 and \#3)  show  \brg\ absorption in the $K$-band and Br10 - 14 absorption in the $H$-band (Fig. \ref{fig:obstars}). Weaker lines of \hei\ are also visible in $H$ (1.70 \micron) and $K$ (2.058 and 2.11 \micron). 
To obtain a spectral type estimate we apply the $K$-band classification scheme for OB stars from  \citet{Hanson96}. Their classification links a K-band spectral type to an optical spectral type. For late O and early B stars, the classification is based on \hei\ (2.11 \micron) and \brg.  For the $H$-band, we used the classification of  \citet{Hanson98} and \citet{Blum97} where the important lines are  \hei\ (1.700 \micron) and Br11 (1.68 \micron).

Table \ref{tab:classification} shows the measured Equivalent Widths (EW)  of the relevant lines in the $H$- and $K$-band spectra of the 3 OB stars. In column 6 the $K$-band spectral type is given, with the corresponding optical spectral type listed in column 7. The brightest star, \#1, has a spectral type between O8V and B1V based on the presence of \hei\ (2.11 \micron) and moderately strong \brg. Star \#2 has a much stronger \brg\ absorption and is therefore classified as B1V - B2.5V. The fainter OB star (\#3) does not have \hei\ (2.11 \micron) in its spectrum and is therefore classified as B3V - B7V. 
 The EWs of the $H$-band lines are in agreement with the  $K$-band spectral type. The only deviating value is the very strong Br11 absorption of \#3, which is about equally strong as the \brg\ line. The strength of Br11 is compatible with that of a supergiant.

The spectral type of star \#1 (O8V-B1V) is compatible with  the optical classification as an O8.5V star by \citet{Heydari88}.  As the optical classification is more reliable than our near-IR classification, we will adopt  a spectral type of O8.5V for \#1 as the main ionizing star for RCW34. This spectral type is in agreement with the Lyman continuum flux
  derived from radio continuum data \citep{Caswell87}, and is also
  compatible with the \hei\ and \brg\ emission of the \HII\ region
  (Sect. 3.3.1). This corresponds to a mass of 20 M$_\odot$ for star \#1  \citep{Martins05}.

\begin{deluxetable}{rlccc}
\tablecaption{The properties of the detected OB stars in RCW 34 \label{tab:OBstars}}
\tablehead{
\colhead{Star}         & 
\colhead{Spectral Type}    & 
\colhead{$K$ (mag)}                    &
\colhead{$J-K$ (mag)} &
\colhead{A$_{V}$ (mag)}
}
\startdata
1 &	O8.5 V   & 8.96 $\pm$ 0.01   &  	0.73 $\pm$ 0.01   	& 4.2 $\pm$ 0.1  \\ 
2 &	B0.5 V  -- B1 V  & 10.35 $\pm$ 0.02 &   0.68 $\pm$ 0.03  	& 3.9 $\pm$ 0.2 \\
3 &	B2 V -- B3 V     & 11.61 $\pm$ 0.03 &   0.75 $\pm$ 0.05  	& 4.1 $\pm$ 0.4 
\enddata

\end{deluxetable}

 The measured color of early type stars is a good measure of the extinction. For the intrinsic colors we used the values given by \citet{Koornneef83}. 
We used the extinction law of \citet{Cardelli89}; the derived extinction towards the three stars is listed in  Table \ref{tab:OBstars} and ranges from A$_{V}$=3.9 to 4.2 mag (A$_{K}$ = 0.42 to 0.45 mag). The error in the A$_{V}$ values reflects both the photometric errors and the uncertainties in ($J$-$K$)$_0$ due to uncertainties in their spectral classification.  These values are in good agreement with those derived from the Br10/\brg\ ratio of the \HII\ spectra (Sect.  3.3.1).

 The spectral types and extinctions of the OB stars, coupled with their
  absolute magnitudes, allow a distance determination.  For the O8.5V
  star \#1, we obtain the absolute magnitude from \citet{Martins06} and \citet{LandoltBornstein}.
This results in  a distance of $2500\pm200$ pc for the O8.5V star.
  The uncertainty reflects the difference in the absolute magnitudes
  between  \citet{Martins06} and \citet{LandoltBornstein}.
     \citet{Hanson97} and \citet{Blum00} provide absolute magnitudes for Zero Age Main Sequence (ZAMS) stars. 
 Their results are about one
  magnitude fainter than corresponding values for the main sequence of
  the Hertzsprung-Russell diagram (HRD). If we adopt the
  ZAMS magnitudes,  we obtain a distance of $1800\pm200$ pc.  The cluster age is
  probably a few Myrs, hence we prefer the main sequence values.
  Thus, we adopt $2500\pm200$ pc as the distance to RCW 34.  This
  distance is slightly closer than the 2.9 kpc derived by
  Heydari-Malayeri (1988) using the same spectral type and optical
  magnitudes.  The difference results from the different absolute magnitudes
  and extinction we apply.

Using the 2.5 kpc distance, we can refine the spectral type
   classification of the two B stars in the cluster. Using the distance modulus, the A$_{V}$ and the observed brightness we derive an absolute $K$-band magnitude of M(K) = -2.1 mag for \#2 and M(K)= -0.8 mag for \#3. With the calibration of  \citet{LandoltBornstein} this results in a B0.5 -- B1 V spectral type for \#2 and a B2-B3V spectral type for \#3 (Table \ref{tab:OBstars}), still compatible with the original near-infrared spectroscopic classification. These spectral types correspond to a mass of 11 and 7 M$_{\sun}$ respectively \citep{LandoltBornstein}.

\subsection{Late-type stars}

\begin{figure*}
\plotone{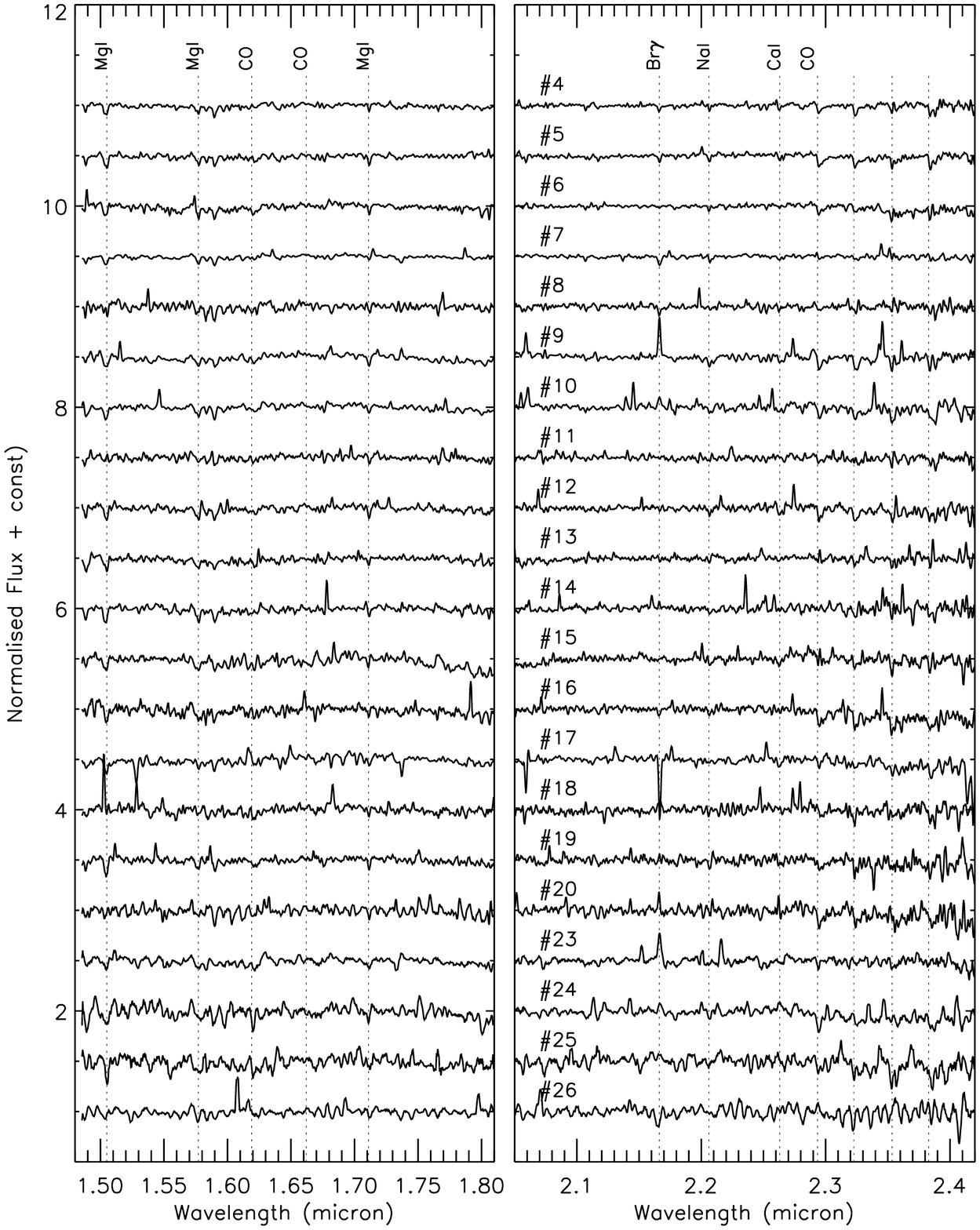}
\caption{Normalized SINFONI $H$- band (left panel) and $K$-band (right panel) spectra of the 21 objects showing late-type stellar spectra. Most of these objects are PMS stars. The vertical dashed lines show the location  of some of the absorption lines used for the classification of the stars. The stars are numbered according to Table \ref{tab:master}.\label{fig:ltstars}}
\end{figure*}

Besides the 3 OB stars we found 21 stars showing absorption lines typical of later spectral type (Fig \ref{fig:ltstars}). The most prominent lines are the CO first overtone absorption bands between 2.29 and 2.45 \micron, and absorption lines of Mg\,{\sc i} and other atomic species in the $H$-band as well as Ca\,{\sc i} and Na\,{\sc i} in the $K$-band. To determine if these stars are cluster members, we need to obtain their  spectral type and luminosity class.

For the  classification of the late-type stars we use the reference spectra of    \citet{Cushing05} and  \citet{Rayner09}.  These spectra are taken with the IRTF telescope\footnote{http://irtfweb.ifa.hawaii.edu/\~\ spex/WebLibrary/index.html}  and cover the spectral types between F and M in all luminosity classes with a resolution between R = 2000 and 5000.
These reference spectra were convolved and rebinned to the resolution of SINFONI.  We corrected for the differences in spectral slope due to the extinction of our SINFONI targets using the extinction law of  \citet{Cardelli89}. The atomic lines, like Mg\,{\sc i} and Na\,{\sc i}, are  used for the temperature determination, while the CO and OH lines serve as a luminosity indicator.

 Comparison of the SINFONI with reference spectra indicates that
   the CO (2.29 \micron) absorption is usually deeper than
   in dwarf reference spectra, but not as deep as in the giant
   spectra.  In the few cases where a luminosity class IV reference
   spectrum was available, the spectrum provided a better match to the
   observed spectrum.   This suggests that our late type stars are low- and intermediate PMS stars; indeed, \citet{Luhman99} and
   \citet{Winston09} find that PMS spectra have a surface gravity intermediate between
   giant and dwarf spectra.
    If the stars were giants, dust veiling could make
   the $K$-band CO lines weaker.  However, the $H$-band CO and OH lines are
   also much weaker, while the atomic lines have the expected EWs. Therefore, this seems to be a surface gravity effect, suggesting a PMS nature of our late type stars.

We used the relation from \citet{Kenyon95} to convert the spectral type into effective temperature. 
This relation applies only for main sequence stars; for PMS stars
   a different relation may hold \citep{Hillenbrand97,Winston09}. \citet{Cohen79} show that the temperature of PMS stars  might be overestimated by values between 500 K (G stars) and 200 K (mid-K). We took this source of error into account when calculating the errors in the effective temperature.  

Table \ref{tab:ltstars} shows the result of the classification for all the late-type stars brighter than K=14 mag.  The spectral types  range from early G to mid K; about half of the stars have a surface gravity indicative of a PMS nature, their luminosity class is marked as "PMS". 
 The stars classified as dwarfs are candidate foreground stars, but
   this does not exclude them as PMS cluster members, as  \citet{Winston09} also find PMS stars with surface gravity very close to the dwarf values.

\begin{deluxetable}{rlcll}

\tablecaption{T$_{\mbox{eff}}$, spectral type, Av and IR-excess of the detected late-type stars brighter than $K$=14 mag.\label{tab:ltstars}} 
\tablehead{
\colhead{Star}			&	
\colhead{T$_{eff}$ (K)}			&	
\colhead{Sp. Type} &
\colhead{Lum. class}& 
\colhead{ IR excess} 
}
\startdata
4	&	5520         $\pm$ 365	& G8 $\pm $ 1& PMS&  -- \\
5	& 	5410 	$\pm$ 365	& G9 $\pm $ 1		& PMS	 & n \\
6	&	5080 	$\pm$ 380	& K1 $\pm $ 1		& PMS	   & + \\
7	&	5860 	$\pm$ 415	& G2 $\pm $ 2		& V	  & + \\
8	&	5680 	$\pm$ 360  	& G6.5 $\pm $ 1		&V	      &--\\
9	&	4900 	$\pm$ 350	&  K2 $\pm $ 1	&PMS 	  &+\\
10	&	5520 	$\pm$ 355	& G8	$\pm $ 1	&PMS&--\\
11	&	5800 	$\pm$ 405  	& G4	$\pm $ 1	&PMS   &+\\
12	&	4900 	$\pm$ 445	& K2	$\pm $ 2	&PMS&n\\
13	&	5680 	$\pm$ 360	& G6.5$\pm $ 1	&PMS      &--\\
14	&	5680 	$\pm$ 360  	&G6.5	$\pm $ 1	&V	   &+\\
15	&	5520 	$\pm$ 355	& G8$\pm $ 1	&V&--\\
16	&	5080 	$\pm$ 380	& K1	$\pm $ 1	&PMS&n\\
17	&	5080 	$\pm$ 380	&  K1 $\pm $1	&PMS	         &n\\
18	&	5080 	$\pm$ 380	& K1	 $\pm $ 1	&V	      &+\\
19	&	4900 	$\pm$ 350	& K2	$\pm $ 1	&V	        &+\\
20	&	4900 	$\pm$ 350	& K2	$\pm $ 1	&PMS	       &n\\
23	&	5520         $\pm$ 355	& G8	$\pm $ 1	&PMS&+\\
24	& 	4800 	$\pm$ 515	& early K	& ?	      &n\\
25	&	4800 $\pm$ 515	& early K	&?         &+\\
26	&	5410	 $\pm$ 365	& G9	 $\pm $ 1	&V	        &n
\enddata
\tablecomments{'+': infrared excess is present, '--' no infrared excess detected, 'n': infrared excess could not be determined  due to bright background emission.}
\end{deluxetable}

With the knowledge of the spectral type, the observed color can be converted into an extinction estimate.  The intrinsic $J-K$ color  for dwarfs and giants can differ as much as 0.4 magnitudes for late K stars  \citep{Koornneef83}. As the late-type stars are PMS stars, we used as the intrinsic $J-K$ color the mean of the color for dwarfs and giants. The difference between the mean value and the dwarf and giant values  is used as the error in the intrinsic color.  The derived extinctions are listed in Table \ref{tab:master}.   Source \#17 does not have a J-band magnitude as the photometry is too
contaminated by source \#1; it is excluded from further analysis.

All the stars with A$_{V} > $  12 mag (\#6, \#8,\#16 and \#24) are located North of the \brg\ emission in the region where the H$_2$ and IRAC dust emission is present. Their extinction is consistent with being embedded in the molecular cloud North of the \HII\ region. The objects situated amidst  the \brg\ emission show  a  visual extinction between 4 and 10 mag, consistent with that measured from the nebular line spectrum (Sect. 3.2.1).

To confirm the PMS nature of the objects we looked for  \brg\ emission from the source. We found two point sources in the continuum subtracted \brg\ line map.  One of the sources corresponds to \#23, a  G8 PMS star. Following \citet{Muzerolle98}, this \brg\ emission could arise from accreting matter.  The other \brg\ emitting point source is \#27 (Table \ref{tab:master}).  This star is rather faint (K= 15.5 $\pm$ 0.1 mag) and the continuum has too low S/N to detect any absorption features. The emission line spectrum also shows \hei\ (2.058 \micron) and the Brackett lines in the $H$-band, identical to the surrounding \HII\ region (Fig \ref{fig:brg_spectrum}).  The \brg\ emission from this object could arise from a disk being photo-evaporated by the main ionizing star. These "proplyds" have been found in Orion \citep{Odell01} and other high-mass star-forming regions like M8 \citep{Stecklum98} and  RCW 38 \citep{DeRose09}. The spatial resolution of our data ($\sim$0.6\arcsec, 1500 AU at 2.5 kpc) did not allow us to resolve the proplyd, which is consistent with the sizes (up to 500 AU) measured for the Orion Proplyds \citep{Odell98}. The observed K-band magnitude and lower-limit of source \#27 does seem consistent with near-infrared observations of the Orion Proplyds \citep{Rost08}.

  In the spectra of e.g. source \#9 and \#17, emission and absorption of \brg\ is detected as well. These lines are artifacts from the \brg\ removal using the median filtering. In the \brg\ line image, no extra emission or absorption is present.

Two objects brighter than $K$=14 mag are unclassified.  These objects
   (\#21 and \#22) have very red spectra and a rather low S/N.  We could
   not identify any features in their spectra. As the spectra are extremely red, they are likely dominated by dust emission and therefore the spectrum of the underlying star might not be visible. Like the late-type stars with high extinction, these featureless spectra are located in the high-extinction area, in the North of the SINFONI field and therefore are most likely embedded stars.

The remaining sources are too faint to obtain a proper classification. In the spectra of  some of these objects, CO absorption at 2.29 \micron\ can still be detected, but the spectra are of too low S/N to detect the atomic lines needed for an  effective temperature classification.  A handful of the objects have a $J-K$ color around 1.0 mag (see Fig. \ref{fig:cmd}), indicating that these might be  late-type foreground stars with very little extinction.  Most of the faint objects are concentrated around $J-K$ = 2.0 mag.  This color is similar to that of the brighter PMS stars around A$_V$ = $5 - 10$ mag. The redder and more deeply embedded objects are not observable as they are too faint to detect in the $J$-band.  The objects with $K > 14$ mag, clustered around $J-K$ = 2 mag, are most likely PMS stars with masses below 1.5 - 2 M$_\sun$. 

\subsection{IR excess}

\begin{figure*}
\plotone{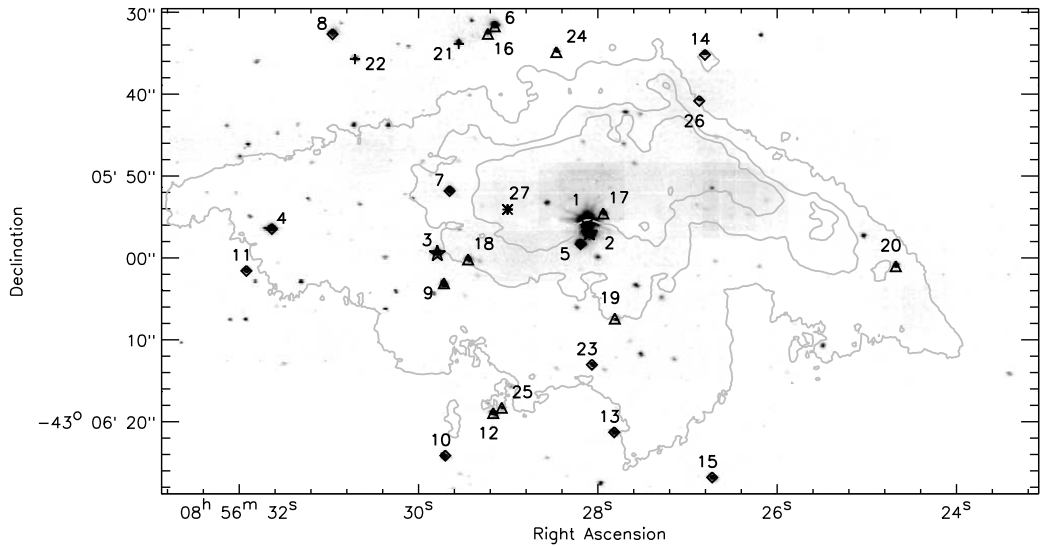}
\caption{Stars brighter than K=14 mag are indicated on  the SINFONI 2.15 micron continuum map. Overplotted in  gray contours is the \brg\ line map. The sources are marked with different symbols representing their spectral types. The star symbols represent the 3 OB stars, the diamonds the G stars and  the triangles the K stars. The featureless stars (\#21 and \#22) correspond to the cross signs. The location of the  proplyd (\#27) is marked with an asterisk. \label{fig:location}}
\end{figure*}

In total, 17 sources (out of 26) have been detected in one or more \emph{Spitzer}/IRAC bands with only three sources detected in all four bands. Most of the sources are only detected at 3.6 and 4.5 \micron. At longer wavelengths, the high and variable background emission makes the detection of point-sources very difficult. The objects detected in all 4 bands are labeled in Fig. \ref{fig:spitzer_ccdiag}. One of the PMS stars (\# 11) is identified as a class II object.  The other 2 objects  (\# 14 and \#21) are classified as class 0/I and will be discussed in Sect. 5.3. 

To determine if the PMS stars have an infrared excess, we construct the SED and compare it with low-resolution Kurucz spectra \citep{Kurucz93} for the 1 to 8 \micron\ range. This allows us to determine the IR excess even for objects detected in  only 1 or 2 IRAC bands.  Column 5 of Table \ref{tab:ltstars} shows that  9 PMS stars show a detectable IR excess (marked as a '+'), while for 5 sources no IR excess is detected (marked as a '--'). The remaining sources are marked with an "n":  no IRAC counterpart could be detected due to blending with other sources or spatial coincidence with the bright, extended arc emission.  The infrared excess is already present at 3.6 \micron\ with an excess of more than a magnitude in some cases.  The detected IR excess of at least 9 of the PMS star candidates reinforces the hypothesis that these stars are young and still surrounded by a circumstellar disk.

The SINFONI spectra, however, suggest that the near-infrared excess of most PMS stars is low, as all these stars have absorption lines in their $K$-band spectra. Should a strong IR excess be present these lines would not have been detected. The $K$-band spectrum of \#11 suggests that some veiling might be present;  the CO was still stronger than that of the best fitting dwarf reference spectrum, but the Na\,{\sc i} and Ca\,{\sc i} lines in the $K$-band were very faint. In the spectra of the two sources where no absorption features have been detected  (\#21 and \#22) a NIR excess is most likely present. These sources also have very red colors and a strong IRAC excess.  

\section{Stellar population}

\subsection{Cluster membership}

\begin{figure*}
\plottwo{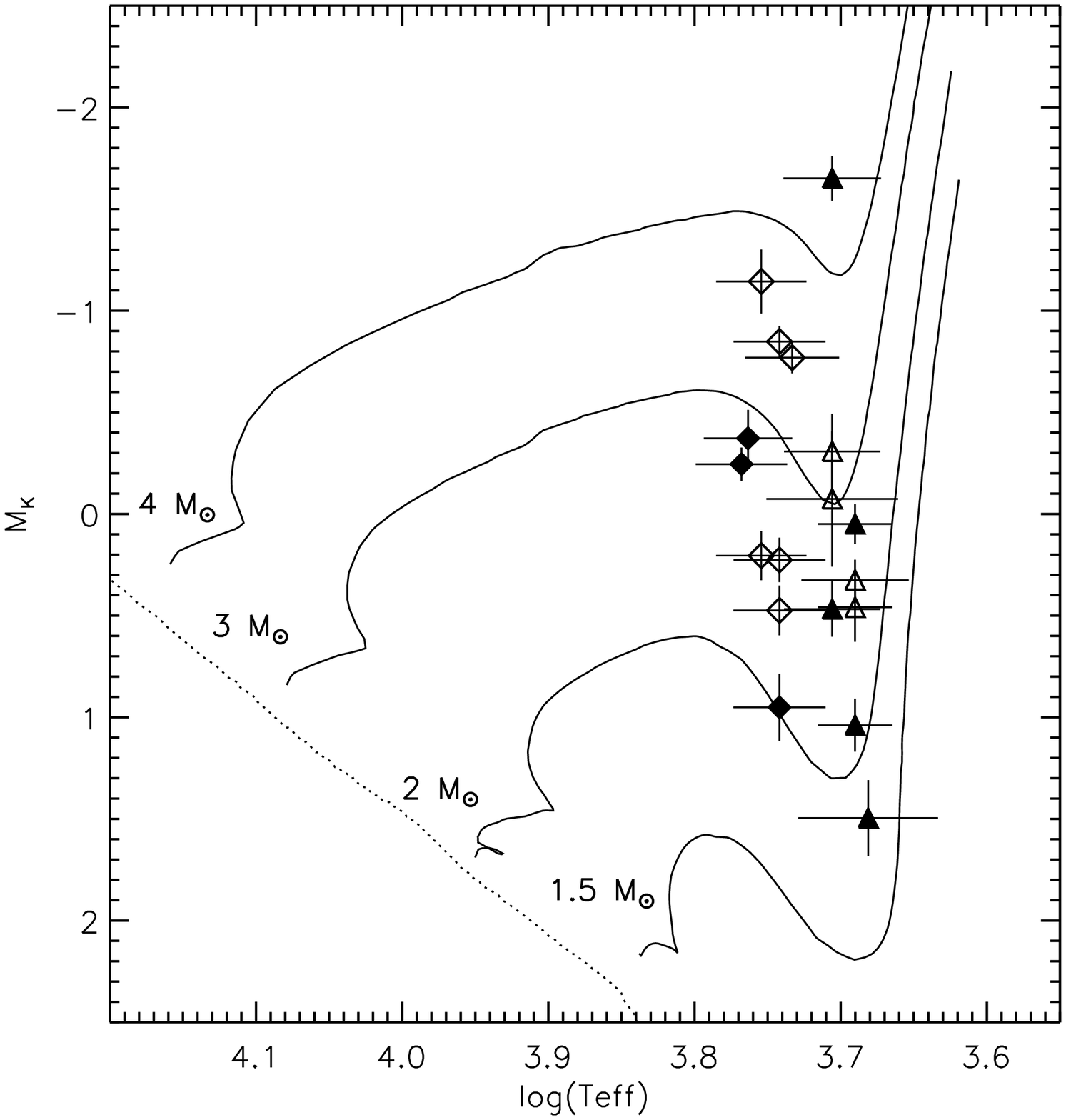}{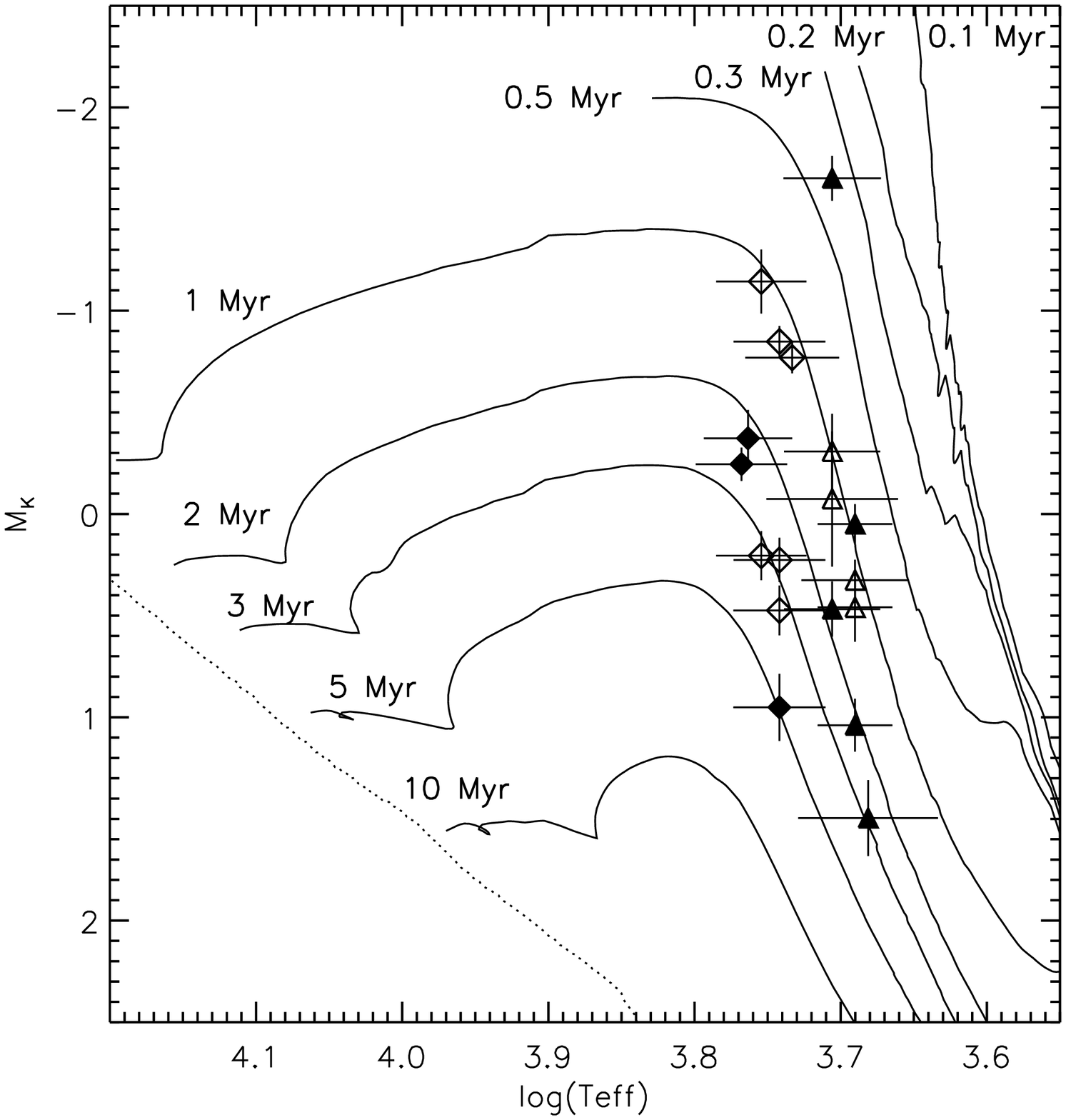}

\caption{\emph{Left}: The extinction corrected $K$ vs log(T$_{\mbox{eff}}$)   HRD. The $K$-band magnitude has been corrected for the distance modulus. The plotting symbols for the stars are the same as in Fig. \ref{fig:cmd}. The filled symbols represent objects with infrared excess detected with IRAC (Table \ref{tab:ltstars}), and the open symbols those without IR excess. Note that in a large fraction of the objects without IR excess, the background emission  is very bright. Over-plotted with a dotted line is the 2 Myr isochrone from \citet{LeJeune01}, and with the solid  lines the pre-main-sequence evolutionary tracks from \citet{Dario09}. \emph{Right}: The same data but overplotted with solid lines showing the
isochrones from \citet{Dario09}. \label{fig:hrd}}
\end{figure*}

To discuss the stellar content in detail, we first need to separate the cluster members from the fore- and background stars.   The O8.5V star (source \#1) ionizes the \HII\ region and is also responsible for the zone of interaction with the molecular cloud to the north.  Analysis of the brightness and spectral type of the B stars (sources \#2 and \#3) shows that they are  located  at the same distance (2.5 kpc) and have a similar extinction as the main ionizing source  and therefore are considered to be cluster members.  

It is less obvious that all  the late-type stars  are cluster members as well. As discussed in Sect.\ 4.2, most of the late-type stars  have a surface gravity typical for PMS stars. Additionally, about half of the stars have an IR excess, supporting the result from the spectroscopy that those stars are PMS stars surrounded by an accretion disk. The extinction derived for these stars is consistent with the extinction derived using the HII region and the H$_{2}$ emission.

Sources \#13 and \#15 show no evidence for a PMS nature nor for an IR excess, but have higher visual extinctions (5.6 and 7 mag, respectively) than observed from the \HII\ emission in that area. Their early spectral types make it difficult to derive a luminosity class because of the CO bands are not present. The high extinction suggests that they are background stars, but then the stars should be (super) giants as dwarfs are too faint to be seen as background stars. 

Two more stars have A$_{V}$ values  inconsistent with their location: sources \#14 and \#26 show very low extinction (0.25 and 1.8 mag, respectively) while they are located towards the high extinction area North of the \HII\ region.  Star \#14 has virtually no extinction and a surface gravity typical for main sequence stars; however, it coincides with an IRAC source having colors of a class 0/I object. Also, on the \brg\ line map some faint \brg\ emission is seen surrounding the star. The fact that this star has no extinction  is inconsistent with its IRAC colors and  a protostar nature. If the star were a protostar, it would be more deeply embedded and have a much larger A$_{V}$. It is most likely a chance alignment of a foreground late G star (as seen in the NIR) with a deeply embedded protostar only seen at mid-IR wavelengths.
Additionally, star \# 26 (A$_{V}$ = 1.8 mag), also located towards the high-extinction area, does not show any PMS characteristic nor an IR excess and also  is most likely a foreground star.

 The locations of the stars in the SINFONI field are shown in Fig. 10.
 The different classes of objects do not cover a preferred area of
 the SINFONI field, but rather are spread across the entire field.

Summarizing, after analysis of their location, reddening and spectral features we conclude that most of the stars are  probably  cluster members, with two exceptions where the stars are probably foreground stars (\#14 and \#26). The nature of sources \#13 and \#15 is unclear, but their high extinction suggests that they are related to the cluster.

\subsection{Stellar content and age}

 To constrain the mass of the PMS stars and the age of
 RCW 34, we construct a HRD to compare the observed parameters with
 PMS evolutionary tracks.  The derived extinctions allow us to  de-redden the K-band magnitude, convert to absolute
 magnitude with the derived temperature, and plot the points in a $K$
 vs. log(T$_{\rm eff}$) diagram (Fig. \ref{fig:hrd}, left).  We exclude the
 foreground stars \#14 and \#26 to enable a comparison with the
 isochrones.  The symbols reflect the different spectral types.
 Filled symbols indicate that a {\it Spitzer}/IRAC excess emission was
 detected.

 Over-plotted as a dotted line is the 2 Myr main sequence isochrone
 taken from \citet{LeJeune01}.  The 2 Myr isochrone is
 chosen to illustrate the main sequence; 1 Myr or 5 Myr isochrones
 would have served equally well.  A 20 M$_{\sun}$ star does not
 evolve significantly off the main sequence for the first 8 Myr of
 its life \citep{LeJeune01}, therefore we adopt 8 Myr as  an upper limit for the age of RCW 34.

The solid  lines in the left panel represent the evolutionary tracks taken from \citet{Dario09} using the models of \citet{Siess00} for 4 different masses; 4, 3, 2, and  1.5 M$_{\sun}$.   Comparison of the location of the PMS stars with these evolutionary tracks yields an approximate mass varying from $\sim$ 3 M$_{\sun}$ for the brightest stars to 1.5 M$_{\sun}$. 

Over-plotted in the right panel  of Fig. \ref{fig:hrd} are the PMS isochrones (solid lines) between 0.1 and 10 Myr. 
The location of the G- and K stars is consistent with a range of ages. The brightest object (\#6) lies between the 0.3 and 0.5 Myr isochrones; however, the majority of the objects span the 1-3 Myr isochrones, with the more massive objects closer to the 1 Myr isochrone. For the less massive objects, the spread in age is larger, most likely due to higher uncertainties in the spectral type determination.  The location of the PMS stars suggests an age of 2 $\pm$ 1 Myr.

 Comparison of the location of the PMS stars in the HRD with those of the Herbig AeBe stars \citep{vandenAncker98} shows that the PMS stars are younger than the Herbig AeBe stars and will evolve from their present G- and K spectral type to late B, A or early F spectral type when they become main sequence stars.  This also explains the fact that no A or F stars are spectroscopically detected, as they are still in the PMS phase at cooler temperatures.

Even though the PMS stars are younger than the Herbig AeBe stars, they have, in contrast to the Herbig stars  \citep[e.g.][]{Eiroa01}, no K-band excess. This could be the result of a different disk structure, but also caused by the fact that the PMS stars are emitting most of their energy in the near-infrared wavelength regime, where the SEDs of  Herbig AeBe stars peak in the optical, thereby outshining the contribution of the circumstellar disk.

The extinction measurements show that the sources North of the \HII\ region are more deeply embedded in the molecular cloud. A younger age for this region is suggested by the presence of several protostars as well as outflows and a dense molecular cloud, which would suggest a younger age for the PMS object in this area as well.
  The PMS stars North of the \HII\ region (\# 6, \#8, \#16, \#24) all have the brightest absolute magnitude in their temperature bin with source \# 6 as an extreme example. Their position in the HRD could suggest that they are the youngest of the stars of given spectral type. However, their high-extinction leads to a selection effect (only the brightest sources are detected). Fainter sources would fall below the limit where SINFONI spectra could be classified. So, if there were an age difference, the current data would not be able to quantify this. Deeper spectroscopic observations to find more of the high-extintion PMS stars North of the \HII\ region would be required to make any definitive statements.

\subsection{Deeply embedded population}

As the SINFONI observations show, large extinction variations are
   present in RCW 34.  The high extinction towards the North means that
   we miss sources that are too deeply embedded to be visible in the K-band.
 In our IRAC catalogue, we found 2 sources with no near-infrared counterpart (\#28 and \#29 in Table \ref{tab:master}). Both sources have IRAC colors of deeply embedded protostars (class 0/I). 
 
Source \#28 is located in the North-East corner of the SINFONI field, only 3\arcsec\ away from one of the outflows  detected in the H$_{2}$ line emission (outflow c,  Sect. 3.2.2). The other embedded source (\#29) is located very close to source \#1, but is not related. \citet{Thomas73} and \citet{Williams77} detected \#29 as a mid-infrared excess towards vdBH 25a.  The IRAC images show, however, that it is a separate  source.  The SED of \# 29 is difficult to determine; at 3.6 \micron\  the contribution of the O8.5V star is still present, and at 5.8 and 8.0 \micron\ this object is the brightest source at MIR wavelengths. To determine a better SED, high-resolution and deeper mid-infrared observations are needed to discriminate between the emission from the protostar and the arc in the north. Additionally, as discussed above,  the IRAC counterpart of source \#14 could be another embedded protostar not seen at NIR wavelengths.

 Outside the SINFONI field, several other protostars are located. Some of these objects seem to have a $K$-band counterpart in the SOFI image of \citet{BikThesis}, suggesting that the objects in our SINFONI field are even more embedded than those located a bit further away from the arc.

\subsection{Stellar population of the surroundings}

IRAC imaging  reveals a large bubble surrounded by extended 8 \micron\ emission. Inside this bubble is a cluster of intermediate and low-mass class II objects. Typical ages for class II objects are of the order of a few Myrs \citep{Kenyon95,White07}, not distinguishable from the age of the stars in the \HII\ region. 

The class 0/I stars, identified mainly close to the bright arc at the northern edge of the \HII\ region, might be significantly younger.  \citet{Whitney03} suggest an age of $\sim 10^5$ years for class I sources; however, \citet{White07} present evidence that the low-mass class I objects in Taurus have a similar age as the class II sources. The location of the class 0/I sources coincides with methanol masers, CS emission, high extinction and outflows, and  therefore are most likely younger than the objects inside the bubble and the \HII\ region.

Comparing the 2MASS $K$-band magnitudes of the class II stars in the bubble  with the $K$-band magnitudes of the PMS stars in the \HII\ region shows that 13 out of 17 class II objects are about a magnitude brighter. The other four are of similar magnitude. Additionally, in a ($J-H$) vs. ($H-K$) color-color plot, most of the class II sources have a near-infrared excess and cover the same region in the color-color diagram as the Herbig Ae stars \citep{Eiroa01}. 

The difference in $K$-band magnitude could be explained with the presence of a $K$-band excess, suggesting that the class II sources are more similar to the Herbig Ae stars and  are therefore more evolved than the PMS stars. An optical spectroscopic classification is needed to confirm this.

The location of the central star of the HII region, near the northern edge of the bubble suggests that this source might not have created the bubble. Also the H$\alpha$ \citep{Parker98} as well as the \brg\ emission peaks at the location of the O star. In the rest of the bubble only  faint H$\alpha$ emission is present, indicating the absence of massive stars inside the bubble. Apart from the class II sources detected in the bubble, one source in the center (Fig. \ref{fig:spitzer_objects}) is detected without infrared excess.  The  magnitude at 3.6 \micron\ (10.0 $\pm$ 0.02 mag) is compatible with a B0V star at the distance of RCW 34.  However, the optical photometry of the source taken from the USNO catalogue \citep{Monet98}, is not compatible with a (reddened) early type star. Therefore we conclude that this star an unrelated, foreground object.

 This raises the question whether other mechanisms could be responsible for the bubble. The scenario that the bubble would be created by a supernova seems unlikely. As the stars in the bubble are only a few Myrs old, the supernova needs to come from an early type O star. Assuming a normal IMF, the other O stars still would have to be visible. 

Massive stars affect their environment by ionizing photons and stellar winds. Estimating the mechanical luminosity (E$_{mech}$ = 1/2 Mv$^2$) needed to blow a bubble with a radius of 2.5 pc inside a medium with a typical molecular cloud density (10$^3$ cm$^{-3}$), expanding at 10 km/s would be on the order of 10$^{42}$ erg/s.  As a comparison, the wind luminosity of a mid-B star is around 10$^{30}$ erg/s \citep{Tan04}.

Assuming  $\sim$10$^{44} $    ionizing photons s$^{-1}$, typical for a mid-B star  and a density of 10$^3$ cm$^{-3}$, the Str\"omgren radius is 0.014 pc.  This is a factor 100 lower than the observed radius of the bubble (2.5 pc).   Assuming that the 20 observed class II stars all emit the same number of ionizing photons and assuming a density of 100 cm$^{-3}$, the Str\"omgren radius would be comparable to that of the measured bubble size. Additionally, the \HII\ will expand into the molecular material due to the higher pressure, making this the most plausible scenario.

Another possibility is that the OB star which created the bubble has been expelled from the center of the cluster due to a binary interaction,  similar to what is suggested for the origin of BN in Orion \citep{Rodriguez05}. The central star of the \HII\ region could be one of the stars from the disrupted binary, with the other star moving in the opposite direction.  With the current data we cannot exclude this scenario, detailed velocity measurements of the star in the HII region could give more insights into this scenario.

\section{Discussion}

The interaction region between the \HII\ region and the molecular cloud North of it raise the suggestion that some form of triggered star formation plays a role here.  After having derived several properties of the \HII\ region, the molecular cloud and the adjacent bubble we can now attempt to reconstruct the way this region has formed. 

RCW 34 is composed of a large bubble  in which intermediate- and low-mass  PMS stars have formed with  no evidence for  a central ionizing source. The ionizing source and its surrounding \HII\ region, however, are located at the edge of the bubble, very close to the molecular cloud. The \HII\ region and the molecular cloud are separated by an ionization front and a PDR traced by H$_{2}$ emission and 5.8 and 8.0 \micron\ emission.
The age of the stars inside the bubble (several Myr) is not known with sufficient accuracy  to detect an age difference with the stars inside the \HII\ region (2 $\pm$ 1 Myrs). The age of the objects forming inside the  molecular cloud, however, is most likely younger, as several tracers of active star formation have been detected. 

Inside the \HII\ region,  a North-South velocity gradient has been detected from H$\alpha$ spectroscopy, resembling that of S88B, a proto-typical champagne flow \citep{Garay98,Stahler05}. The detection of the champagne flow, as well as the presence of the ionization front and PDR in the North, suggest that the \HII\ region is formed at the edge of the molecular cloud. Due to the lower-density in the bubble, the HII region can expand into the bubble creating the observed velocity gradient.

We already can exclude one of the triggered star-formation scenarios. The collect and collapse scenario, where the star-forming clumps are formed by swept-up material is very likely not the scenario that plays a role here. Firstly, only at the northern edge of the bubble is a molecular clump present, while more clumps around the bubble are expected \citep{Deharveng05}. Secondly, it is  very unlikely that the energy output of a cluster of only intermediate mass stars would be enough to sweep up a dense molecular clump of more than 1000 M$_\sun$ as detected in CS emission.

Also unlikely is that the formation of the \HII\ region and the O star caused both the bubble and the star formation in the Molecular Cloud. The difference in evolutionary stage (class II stars in the bubble and class 0/I stars in the molecular cloud) is difficult to explain by the triggering due to the \HII\ region alone.

The location of the bubble, \HII\ region and molecular cloud suggest a sequence of events. First,  the cluster of stars formed which created the bubble. After that, the HII region formed at the
  border of the bubble and the larger molecular cloud; the latter
  currently forming stars. The location of the \HII\ region with respect to the molecular cloud as well as the presence of the PDR shows that the O star is influencing the molecular cloud and probably also inducing the current star formation. 

The question remains whether  the formation of the bubble triggered the other star-formation events.  A mechanism whereby  the  energy output of the stars inside the bubble would be pushing a pre-existing core to collapse could be at work here. If a molecular core present at the edge of the molecular cloud were in equilibrium, a small push from the bubble could have made the core collapse and enable the formation of the OB stars. 

Another alternative would be that the triggering mechanism is external and that a wave of star formation is propagating through the molecular cloud resulting in the formation of the different sites of star formation.

To determine which scenario is the correct one, a better age determination of the class II sources by means of optical or near-infrared spectroscopy would be needed. A triggering scenario would result in an age difference between the stars in the bubble and the \HII\ region.

\section{Conclusions}

In this paper we have presented SINFONI integral field spectroscopy of the \HII\ region RCW 34 as well as \emph{Spitzer}/IRAC imaging of the surrounding area.  
The results can be summarized as follows.

1. RCW 34 consists of 3 separate regions, a large bubble outlined by \emph{Spitzer}/IRAC emission with an \HII\ region at the northern edge of the bubble and a dense molecular cloud just North of the \HII\ region.   IRAC photometry of the sources inside the bubble reveals a cluster of intermediate class II sources and shows that no OB star is located near the center of the bubble. The class 0/I objects are mostly detected towards the dense molecular cloud in the north, suggesting active star formation present in the molecular cloud.

2. SINFONI spectroscopy of the \HII\ region reveals an extinction gradient from North (A$_{V}$ =15.9 mag)  to South (A$_{V}$ = 1.5 mag), supportive of the champagne flow hypothesis measured in the H$\alpha$ velocity profile. The region between the \HII\ region and the molecular cloud is identified as a PDR, where the H$_{2}$ emission originates from UV fluorescence. Additionally, two outflows are identified based on their H$_{2}$ emission.

3. SINFONI spectroscopy identifies the 3 brightest stars as ionizing sources of the \HII\ region. The main ionizing source, \#1, is of spectral type O8.5V while the two other stars are B0.5V and B3V stars.  This results in a revised distance estimate of 2.5 $\pm$ 0.2 kpc for RCW34.   Additionally, we detected 21 G and K stars of which two stars are foreground objects and 19 objects are PMS cluster members. With a mass between 2 and 3 M$_{\sun}$, these objects are still contracting to the main sequence and are the precursors of the A and F stars. Combining the location of those objects in a HRD with PMS evolutionary models results in an age estimate of 2  $\pm$ 1 Myr for RCW 34.

4.  In this scenario, the cluster stars created the bubble, which
   then collapsed a pre-existing dense core at the edge of the
   molecular cloud.  The resulting \HII\ region would encounter
   the density gradient at the edge of the cloud, and thus develop
   as a champagne flow.  Whether the bubble was the original trigger,
   or if there was some other, external trigger, cannot be answered
   with the current data; a detailed spectroscopic study of the
   bubble stars will be required.

 This study shows that the interaction between the young massive stars and the surrounding molecular cloud could not only lead to the destruction of the cloud, but also to the formation of a younger generation of stars. In the majority of our regions we have identified multiple generations of star formation. Several papers are in preparation to discuss the remainder of the sample of 10 clusters observed with SINFONI  (e.g. Wang et al, in prep, Vasyunina et al. in prep.)  and more final conclusions will be presented as a final paper discussing the survey.

\acknowledgments
The authors thank F. Eisenhauer for providing the data reduction software SPRED,  A. Modigliani for help in the data reduction,  R. Davies for providing his routines to remove the OH line residuals and N. Da Rio for providing the isochrones and evolutionary tracks as well as L. Decin for discussions on the classification of the PMS stars. The authors thank the Paranal staff for carrying out the observations. 

EP is funded by the Spanish MEC under the Consolider-Ingenio 2010 Program grant CSD2006-00070: First Science with the GTC, and under grants AYA2007-67456-C02-02/AYA2008-06166-C03-02. HB and AS acknowledge financial support by the Emmy Noether Program of the Deutsche Forschungsgemeinschaft (DFG, grants BE2578, STO 496/3). MBNK was supported by the Peter and Patricia Gruber Foundation and by PPARC/STFC under grant number PP/D002036/1.
This work is based on observations made with the \emph{Spitzer Space Telescope}, which is operated 
by the Jet Propulsion Laboratory, California Institute of Technology under a contract with the National Aeronautics and 
Space Administration ( NASA). NSO/Kitt Peak FTS data used here were produced by NSF/NOAO.

{\it Facilities:} \facility{VLT:Yepun (SINFONI)}, \facility{Spitzer (IRAC)}


\newpage

 \begin{landscape}

\begin{deluxetable*}{rccccccccrrc}
\tablecaption{Photomeric and spectroscopic properties of the sources located in the SINFONI field of view. \label{tab:master}}
\tablehead{
\colhead{Star} &
\colhead{RA (2000)} &
\colhead{DEC (2000)} &
\colhead{$K$ } &
\colhead{$J-K$} &
\colhead{[3.6]} &
\colhead{[4.5]} &
\colhead{[5.8]} &
\colhead{[8.0]} &
\colhead{Sp. Type} &
\colhead{A$_{V}$} &
\colhead{Comments} \\
\colhead{}&
\colhead{(h m s)} & 
\colhead{($^\circ$\ \arcmin\  \arcsec)} &
\colhead{(mag)}&
\colhead{(mag)}&
\colhead{(mag)}&
\colhead{(mag)}&
\colhead{(mag)}&
\colhead{(mag)}&
\colhead{}&
\colhead{(mag)}&
\colhead{}
}
\startdata
 1 &  08:56:28.1  & -43:05:55.6    &  8.96  $\pm$ 0.01  & 0.73  $\pm$ 0.01  & \nodata   & \nodata  & \nodata  & \nodata   & O8.5V& 4.8 $\pm$ 0.1 & vdBH 25a \\                                                   
2  &  08:56:28.1  & -43:05:57.1    &  10.35  $\pm$ 0.02  & 0.69  $\pm$ 0.03  & \nodata   & \nodata  & \nodata  & \nodata  & B0.5-B1V& 4.1 $\pm$ 0.2\\                                                           
3  &  08:56:29.8  & -43:05:59.4    &  11.61  $\pm$ 0.03  & 0.76  $\pm$ 0.05  &  10.81  $\pm$ 0.03  &  10.37  $\pm$ 0.06   & \nodata  & \nodata & B2-B3V& 4.3 $\pm$ 0.4\\                                        
4  &  08:56:31.7  & -43:05:56.3    &  11.66  $\pm$ 0.04  & 1.56  $\pm$ 0.06  &  11.26  $\pm$ 0.03  &  11.01  $\pm$ 0.06   & \nodata  & \nodata  &G8$\pm$1 PMS & 4.7 $\pm$ 0.6\\                                 
5  &  08:56:28.2  & -43:05:58.3    &  11.73  $\pm$ 0.04 & 1.54  $\pm$ 0.06  & \nodata  & \nodata  & \nodata  & \nodata  &G9$\pm$1 PMS & 4.6 $\pm$ 0.6\    \\                                                  
6  &  08:56:29.1  & -43:05:31.6    &  12.01  $\pm$ 0.04 & 3.53  $\pm$ 0.15  & 9.67  $\pm$ 0.02  & \nodata   & 8.51  $\pm$ 0.16  & \nodata  & K1$\pm$1 PMS   & 14.2 $\pm$ 0.9  \\                                
7 &  08:56:29.6  & -43:05:51.9    &  12.23  $\pm$ 0.05  & 1.42  $\pm$ 0.07  &  10.84  $\pm$ 0.17  & \nodata  & \nodata  & \nodata    & G2$\pm$ 2 V & 4.4 $\pm$ 0.6 \\                                           
8  &  08:56:31.0  & -43:05:32.4    &  12.61  $\pm$ 0.05  & 3.6  $\pm$ 0.2  &  11.61  $\pm$ 0.07    & \nodata  & \nodata  & \nodata  & G6.5 $\pm$ 1 V  &15.0 $\pm$ 1.3     \\                                    
9  &  08:56:29.7  & -43:06:03.2    &  12.68  $\pm$ 0.06  & 1.93  $\pm$ 0.11  &  10.88  $\pm$ 0.03  &  10.49  $\pm$ 0.04  & 9.7  $\pm$ 0.10  & \nodata  & K2 $\pm$1 PMS/V &5.7 $\pm$ 0.7  \\                       
10 &  08:56:29.7  & -43:06:24.3    &  12.83  $\pm$ 0.06  & 1.72  $\pm$ 0.11  &  12.17  $\pm$ 0.03  &  11.88 $\pm$ 0.05  & \nodata  & \nodata  & G8 $\pm$1 PMS  & 5.5 $\pm$ 0.8  \\                             
11 &  08:56:31.9  & -43:06:01.5    &  12.87  $\pm$ 0.06  & 2.71  $\pm$ 0.16  &  11.00  $\pm$ 0.02  &  10.36 $\pm$ 0.02  & 9.76  $\pm$ 0.04  & 8.92  $\pm$ 0.08  &G4 $\pm$1 PMS/V  & 10.8 $\pm$ 1.1  & class II\\          
12&  08:56:29.1  & -43:06:19.1    &  12.88  $\pm$ 0.06  & 1.80  $\pm$ 0.11 & \nodata  & \nodata  & \nodata  & \nodata  &  K2 $\pm$2 PMS/V  &5.0 $\pm$ 0.7   \\                                                  
13&  08:56:27.8  & -43:06:21.4    &  12.99  $\pm$ 0.06  & 1.99  $\pm$ 0.13 &  12.17  $\pm$ 0.04  &  11.67  $\pm$ 0.07  & \nodata  & \nodata  & G6.5 $\pm$ 1 V  & 7.0 $\pm$ 0.9   \\                               
14 &  08:56:26.8  & -43:05:35.2    &  13.01  $\pm$ 0.07  & 0.71  $\pm$ 0.09  &  10.51 $\pm$ 0.12  & 9.89 $\pm$ 0.10  & 7.10 $\pm$ 0.11  & 5.21  $\pm$ 0.11 & G6.5 $\pm$ 1 V & 0.3 $\pm$ 0.7  & vdBH 25a C \\    
15&  08:56:26.7  & -43:06:27.0    &  13.09  $\pm$ 0.07  & 1.73  $\pm$ 0.12  &  12.38  $\pm$ 0.04  &  12.00  $\pm$ 0.07  & \nodata  & \nodata  & G8 $\pm$ 1 V & 5.6 $\pm$ 0.9   \\                               
16 &  08:56:29.2  & -43:05:32.5    &  13.22  $\pm$ 0.07  & 3.32  $\pm$ 0.25  & \nodata  & \nodata  & \nodata  & \nodata  & K1 $\pm$ 1 PMS  & 13.1 $\pm$ 1.5    \\                                             
17&  08:56:28.0  & -43:05:55.2    &  13.23  $\pm$ 0.07  &  \nodata  $\pm$ 0.08  & \nodata  & \nodata  & \nodata  & \nodata  & K1$\pm$ 1 PMS/V  & \nodata   \\                                                         
18 &  08:56:29.4  & -43:06:00.3    &  13.29  $\pm$ 0.07  & 2.17  $\pm$ 0.16  &  10.90  $\pm$ 0.03  &  10.21 $\pm$ 0.04  & 9.42  $\pm$ 0.10  & \nodata  & K1 $\pm$ 1 V   & 7.2 $\pm$ 1.0  \\                     
19 &  08:56:27.8  & -43:06:07.5    &  13.47  $\pm$ 0.08  & 1.60  $\pm$ 0.14  &  12.01  $\pm$ 0.10  &  11.06  $\pm$ 0.11  & \nodata  & \nodata  & K2 $\pm$ 1  V & 4.0 $\pm$ 0.9 \\                                 
20&  08:56:24.6  & -43:06:01.1    &  13.54  $\pm$ 0.08  & 2.66  $\pm$ 0.21  & \nodata  & \nodata  & \nodata  & \nodata  &  K2 $\pm$ 1 PMS & 9.4 $\pm$ 1.3   \\                                                  
21 &  08:56:29.6  & -43:05:34.1    &  13.70  $\pm$ 0.09  &  $>$ 4.20  &  10.01 $\pm$ 0.03  & 9.03  $\pm$ 0.03  & 8.23  $\pm$ 0.08  & 6.96  $\pm$ 0.13   & featureless  & class 0/I    \\                           
22 &  08:56:30.7  & -43:05:35.4    &  13.76  $\pm$ 0.09  &  $>$ 4.14   &  11.25  $\pm$ 0.07  &  10.53  $\pm$ 0.06  & \nodata  & \nodata  & featureless  & \nodata & \nodata  \\                                          
23 &  08:56:28.0  & -43:06:13.2    &  13.76  $\pm$ 0.09  & 2.05  $\pm$ 0.19  &  12.40  $\pm$ 0.11  &  11.66  $\pm$ 0.15  & \nodata  & \nodata  & G8 $\pm$ 1 PMS/V & 7.2 $\pm$ 1.2 & \brg\ emission \\                            
24 &  08:56:28.4  & -43:05:34.9    &  13.80  $\pm$ 0.09  & 4.01  $\pm$ 0.46  & \nodata  & \nodata  & \nodata  & \nodata  & early K& 15.9 $\pm$ 2.8   \\                                               
25&  08:56:29.1  & -43:06:18.5    &  13.88  $\pm$ 0.10  & 1.57  $\pm$ 0.17  &  11.75  $\pm$ 0.04  &  11.27  $\pm$ 0.06  &  10.8  $\pm$ 0.15  & \nodata  & early K & 3.6 $\pm$ 1.4   \\                 
26 &  08:56:26.8  & -43:05:40.8    &  13.93  $\pm$ 0.10  & 1.06  $\pm$ 0.15  & \nodata  & \nodata  & \nodata  & \nodata  & G9 $\pm$ 1 V  & 1.8 $\pm$ 1.0 & vdBH 25q   \\                                        
27 &   08:56:29.0   & -43:05:54.0  & 15.5  $\pm$ 0.1        &   $>$2.3       & \nodata  & \nodata  & \nodata  & \nodata  & \nodata  & \nodata  & Proplyd    \\  
28 & 08:56:32.5 & -43:05:34.69   & \nodata                       & \nodata        & 11.69 $\pm$ 0.03  & 9.77 $\pm$ 0.02  & 8.16 $\pm$ 0.02  &6.00 $\pm$ 0.03  & \nodata  & \nodata  & class 0/I \\
29 & 08:56:28.0 & -43:05:53.76   & \nodata		&  \nodata             & 8.24 $\pm$ 0.01  &7.57 $\pm$ 0.02  & 6.43 $\pm$ 0.03  & 4.49 $\pm$ 0.04  & \nodata  & \nodata  & class 0/I
 \enddata
\end{deluxetable*}

         \clearpage
       \end{landscape}

\end{document}